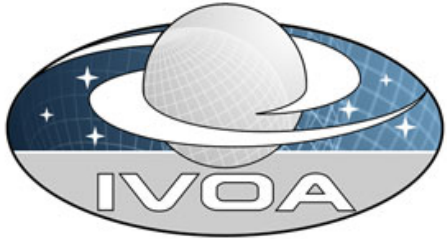

# VODataService: a VOResource Schema Extension for Describing Collections and Services
# Version 1.1

## IVOA Recommendation 02 December 2010

**This version:**
http://www.ivoa.net/Documents/VODataService/20101202
**Latest version:**
http://www.ivoa.net/Documents/VODataService
**Previous versions:**
PR: http://www.ivoa.net/Documents/VODataService/20100916
PR: http://www.ivoa.net/Documents/VODataService/20100914
PR: http://www.ivoa.net/Documents/VODataService/20100412
PR: http://www.ivoa.net/Documents/VODataService/20090903
WD: http://www.ivoa.net/Documents/WD/ReR/VODataService-20090508.html

**Authors:**
Raymond Plante, Editor
Aurélien Stébé
Kevin Benson
Patrick Dowler
Matthew Graham
Gretchen Greene
Paul Harrison
Gerard Lemson
Tony Linde
Guy Rixon
and the IVOA Registry Working Group.

---

## Abstract

VODataService refers to an XML encoding standard for a specialized extension of the IVOA Resource Metadata that is useful for describing data collections and the services that access them. It is defined as an extension of the core resource metadata encoding standard known as VOResource [Plante et al. 2008] using XML Schema. The specialized resource types defined by the VODataService schema allow one to describe how the data underlying the resource cover the sky as well as cover frequency and time. This coverage description leverages heavily the Space-Time Coordinates (STC) standard schema [Rots 2007]. VODataService also enables detailed descriptions of tables that includes information useful to the discovery of tabular data. It is intended that the VODataService data types will be particularly useful in describing services that support standard IVOA service protocols.

## Status of this document

This document has been produced by the IVOA Registry Working Group. It has been reviewed by IVOA Members and other interested parties, and has been endorsed by the IVOA Executive Committee as an IVOA Recommendation as 01 Oct 2010. It is a stable document and may be used as reference material or cited as a normative reference from another document. IVOA's role in making the Recommendation is to draw attention to the specification and to promote its widespread deployment. This enhances the functionality and interoperability inside the Astronomical Community.

A list of current IVOA Recommendations and other technical documents can be found at http://www.ivoa.net/Documents/.

## Acknowledgements

This document has been developed with support from the National Science Foundation's Information Technology Research Program under Cooperative Agreement AST0122449 with The Johns Hopkins University, from the UK Particle Physics and Astronomy Research Council (PPARC), from the European Commission's (EC) Sixth Framework Programme via the Optical Infrared Coordination Network (OPTICON), and from EC's Seventh Framework Programme via its eInfrastructure Science Repositories initiative.

### Conformance-related definitions

The words "MUST", "SHALL", "SHOULD", "MAY", "RECOMMENDED", and "OPTIONAL" (in upper or lower case) used in this document are to be interpreted as described in IETF standard, RFC 2119 [RFC 2119].

The **Virtual Observatory (VO)** is general term for a collection of federated resources that can be used to conduct astronomical research, education, and outreach. The **International Virtual Observatory Alliance (IVOA)** is a global collaboration of separately funded projects to develop standards and infrastructure that enable VO applications.

XML document **validation** is a software process that checks that an XML document is not only well-formed XML but also conforms to the syntax rules defined by the applicable schema. Typically, when the schema is defined by one or more XML Schema [schema] documents (see next section), validation refers to checking for conformance to the syntax described in those Schema documents. This document describes additional syntax constraints that cannot be enforced solely by the rules of XML Schema; thus, in this document, use of the term validation includes the extra checks that go beyond common Schema-aware parsers which ensure conformance with this document.

### Syntax Notation Using XML Schema

The eXtensible Markup Language, or XML, is a document syntax for marking textual information with named tags and is defined by the World Wide Web Consortium (W3C) Recommendation, XML 1.0 [XML]. The set of XML tag names and the syntax rules for their use is referred to as the document schema. One way to formally define a schema for XML documents is using the W3C standard known as XML Schema [schema].

This document defines the VOResource schema using XML Schema. The full Schema document is listed in Appendix A. Parts of the schema appear within the main sections of this document; however, documentation nodes have been left out for the sake of brevity.

References to specific elements and types defined in the VOResource schema include the namespaces prefix, `vr`, as in `vr:Resource` (a type defined in the VOResource schema). References to specific elements and types defined in the VODataService extension schema include the namespaces prefix, `vs`, as in `vs:DataCollection` (a type defined in the VODataService schema). Use of the `vs` prefix in compliant instance documents is strongly recommended, particularly in the applications that involve IVOA Registries (see [RI], section 3.1.2). Elsewhere, the use is not required.

# Contents

---

# 1. Introduction

The VOResource standard [VOR] provides a means of encoding IVOA Resource Metadata [RM] in XML. VOResource uses XML Schema [schema] to define most of the XML syntax rules (while a few of the syntax rules are outside the scope of Schema). VOResource also describes mechanisms for creating extensions to the core VOResource metadata. This allows for the standardization of new metadata for describing specialized kinds of resources in a modular way without deprecating the core schema or other extensions. This document defines one such extension referred to as **VODataService**.

## 1.1. The Role in the IVOA Architecture

The IVOA Architecture [Arch] provides a high-level view of how IVOA standards work together to connect users and applications with providers of data and services, as depicted in the diagram in Fig. 1.

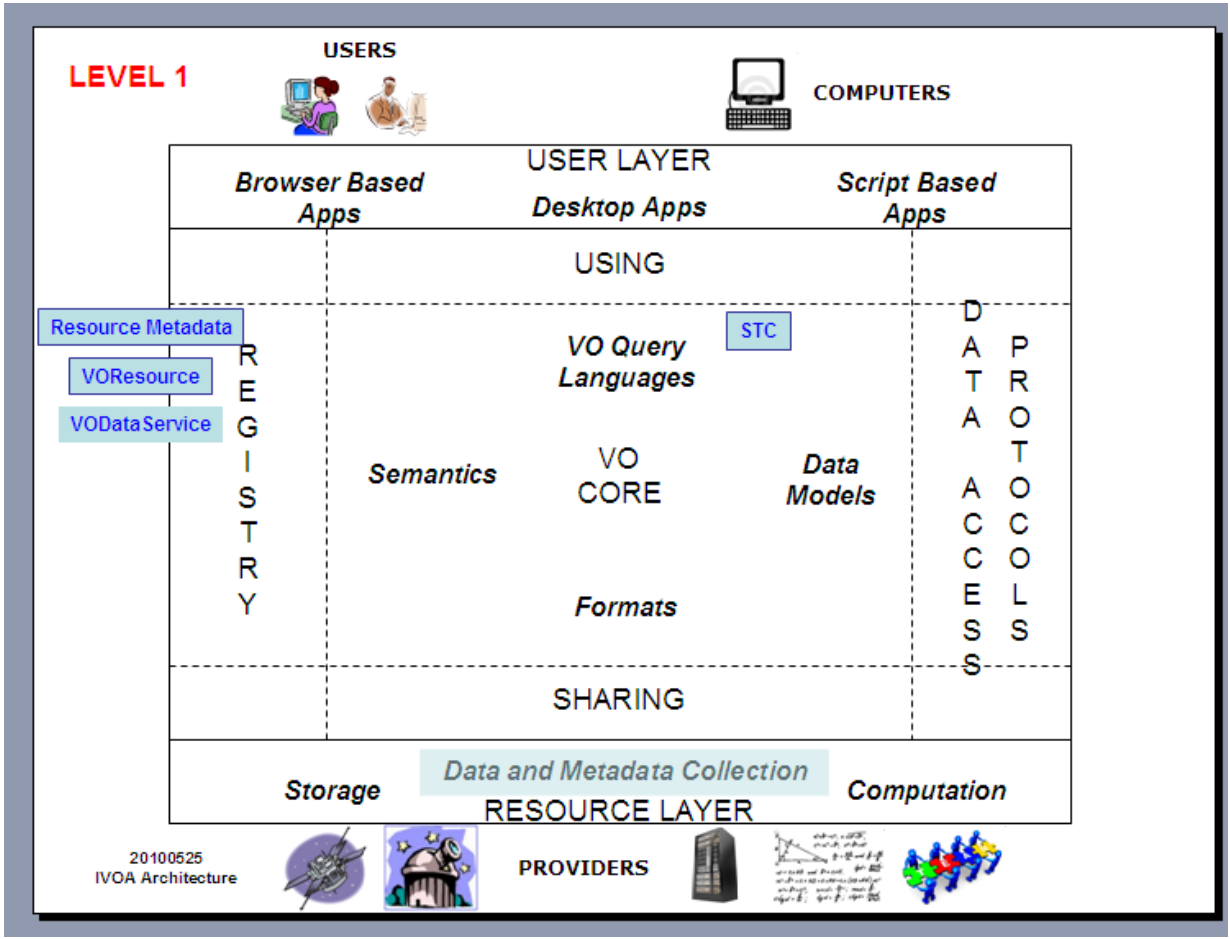

**Figure 1. VODataService in the IVOA Architecture.** The Registry enables applications in the User Layer to discover archives and services in the Resource Layer. The metadata and data model standards (in blue text and boxes) gives structure to the information that enables that discovery.

In this architecture, users can leverage a variety of tools (from the User Layer) to discover archives and services of interest (represented in the Resource Layer); registries provide the means for this discovery. A registry is a repository of descriptions of resources that can be searched based on the metadata in those descriptions. The Resource Metadata standard [RM] defines the core concepts used in the resource descriptions, and the VOResource standard [VOR] defines the XML format. As an extension of VOResource, the VODataService standard, defined in this document, specifically supports descriptions of data collections and services.

## 1.2. Purpose

The purpose of this extension is to define common XML Schema types--particularly new resource types--that are useful for describing data collections and services that access data. In particular, it allows one to describe the data's *coverage*: the parts of the sky with which the data are associated and the time and frequency ranges that were observed or modeled to create the data. It also allows one to describe tables in detail. In particular, one can describe each of the columns of a table--providing, for example, its name, type, UCD [UCD], and textual description. When this metadata is part of a resource description in a registry [VOR], it becomes possible to discover tables that contain particular kinds of data.

It is intended that VODataService will be central to describing services that support standard IVOA data access layer protocols such as Simple Image Access [SIA] and Simple Cone Search [SCS]. While other VOResource extensions would define the protocol-specific metadata (encapsulated as a standard *capability* [VOR]), the general service resource description would share the common data concepts such as coverage and tabular data. Note, however, that a service described using the VODataService schema need not support any standard protocols. With the VODataService extension schema plus the core VOResource schema, it is possible to describe a custom service interface that accesses data.

As a legal extension of VOResource [VOR], the use of VODataService is subject to the rules and recommendations for XML [xml], XML Schema [schema], and VOResource itself.

## 2. The VODataService Data Model

The VODataService extension in general enables the description of two types of resources: data collections and services that access data. Here's an example of a VOResource document (abbreviated for the purposes of illustration) that describes a service from the NASA Extragalactic Database (NED) that provides measured redshifts for a given object.

---

**Example**

A description of a service returning tabular data, catalogservice.xml

```
       <?xml version="1.0" encoding="UTF-8"?>
       <ri:Resource xmlns=""
2          xsi:type="vs:CatalogService" status="active"
              updated="2008-04-29T14:51:54" created="2005-10-14T01:46:00"
              xmlns:ri="http://www.ivoa.net/xml/RegistryInterface/v1.0"
              xmlns:vr="http://www.ivoa.net/xml/VOResource/v1.0"
1             xmlns:vs="http://www.ivoa.net/xml/VODataService/v1.1"
1             xmlns:stc="http://www.ivoa.net/xml/STC/stc-v1.30.xsd"
1             xmlns:xlink="http://www.w3.org/1999/xlink"
              xmlns:xsi="http://www.w3.org/2001/XMLSchema-instance"
3             xsi:schemaLocation="http://www.ivoa.net/xml/VOResource/v1.0
3                       http://www.ivoa.net/xml/VOResource/v1.0
3                       http://www.ivoa.net/xml/VODataService/v1.1
3                       http://www.ivoa.net/xml/VODataService/v1.1
3                       http://www.ivoa.net/xml/STC/stc-v1.30.xsd
3                       http://www.ivoa.net/xml/STC/stc-v1.30.xsd">

4      <title>The NASA/IPAC Extragalactic Database</title>
4      <shortName>NED_redshift</shortName>
4      <identifier>ivo://ned.ipac/Redshift_By_Object_Name</identifier>
4      <curation>
4        <publisher>The NASA/IPAC Extragalactic Database</publisher>
4        <contact>
4          <name>Olga Pevunova</name>
4          <email>contact@datacenter.edu</email>
4        </contact>
4      </curation>
4      <content>
4        <subject>redshift</subject>
4        <subject>galaxies</subject>
4        <description>
4          NED is built around a master list of extragalactic objects for
```

```
4        which cross-identifications of names have been established,
4        accurate positions and redshifts entered to the extent possible,
4        and some basic data collected. This service will return recorded
4        redshifts for a given object.
4      </description>
4      <referenceURL>http://nedwww.ipac.caltech.edu/help/data_help.html#zdat</referenceURL>
4      <type>BasicData</type>
4      <contentLevel>Research</contentLevel>
4    </content>

4    <capability>
5      <interface xsi:type="vs:ParamHTTP">
5        <accessURL use="base">
5          http://nedwww.ipac.caltech.edu/cgi-bin/nph-datasearch?search_type=Redshifts&
5        </accessURL>
5        <queryType>GET</queryType>
5        <resultType>application/xml+votable</resultType>
5        <param use="required">
5          <name>objname</name>
5          <description>Name of object</description>
5          <dataType>string</dataType>
5        </param>
5        <param use="required">
5          <name>of</name>
5          <description>Output format parameter, must be "xml_main" for VOTable output.</description>
5          <dataType>string</dataType>
5        </param>
5      </interface>
4    </capability>

6    <coverage>
6      <stc:STCResourceProfile>
6        <stc:AstroCoordSystem xlink:type="simple"
6                              xlink:href="ivo://STClib/CoordSys#UTC-FK5-TOPO"
6                              id="UTC-FK5-TOPO"/>
6        <stc:AstroCoordArea coord_system_id="UTC-FK5-TOPO">
6          <stc:AllSky/>
6        </stc:AstroCoordArea>
6      </stc:STCResourceProfile>

6      <waveband>Radio</waveband>
6      <waveband>Optical</waveband>
6    </coverage>

7    <tableset>
7      <schema>
7        <name>default</name>
7        <table type="output">
7          <name>default</name>
7          <column>
7            <name>No.</name>
7            <description>
7              A sequential data-point number applicable to this list only.
7            </description>
7            <ucd>meta.number</ucd>
7            <dataType xsi:type="vs:VOTableType">int</dataType>
7          </column>
7          <column>
7            <name>Name in Publication</name>
7            <description>
7              The object's name in NED's standard format, of the object to which the data apply.
7            </description>
7            <ucd>meta.id;name</ucd>
7            <dataType xsi:type="vs:VOTableType">string</dataType>
7          </column>
7          <column>
```

```
7            <name>Published Velocity</name>
7            <description>
7              The radial velocity , derived from derived from the shift of some spectral feature, in km/sec
7            </description>
7            <unit>km/sec</unit>
7            <ucd>src.spect.dopplerVeloc</ucd>
7            <dataType xsi:type="vs:VOTableType">int</dataType>
7          </column>
7        </table>
7      </schema>
7    </tableset>
  </ri:Resource>
```

This example illustrates some of the features of the VODataService extension:

1. the extra namespaces associated with VODataService metadata; if STC coverage information [STC] is not included, then only the VODataService namespaced is needed.
2. the specific type of resource indicated by the value of the `xsi:type` attribute; in this case `vs:CatalogService` indicates that this is describing a service that accesses tabular data.
3. the location of the VOResource-related schema documents used by this description,
4. the core VOResource metadata,
5. an interface described by the VODataService interface type, `vs:ParamHTTP`; this type can indicate input arguments it supports.
6. a description of the coverage, including an STC description plus waveband keywords.
7. a description of the table that is returned by the service.

## 2.1. The Schema Namespace and Location

The namespace associated with VODataService extensions is "http://www.ivoa.net/xml/VODataService/v1.1". Just like the namespace URI for the VOResource schema, the VODataService namespace URI can be interpreted as a URL. Resolving it will return the XML Schema document (given in Appendix A) that defines the VODataService schema.

Authors of VOResource instance documents may choose to provide a location for the VOResource XML Schema document and its extensions using the `xsi:schemaLocation` attribute. While the choice of the location value is the choice of the author, this specification recommends using the VODataService namespace URI as its location URL (as illustrated in the example above), as in,

```
xsi:schemaLocation="http://www.ivoa.net/xml/VODataService/v1.1
                    http://www.ivoa.net/xml/VODataService/v1.1"
```

> **Note:**
>
> The IVOA Registry Interface standard [RI] actually *requires* that the VOResource records it shares with other registries provide location URLs via `xsi:schemaLocation` for the VOResource schema and all legal extension schemas that are used in the records. This rule would apply to the VODataService schema.

The prefix, `vs`, is used by convention as the prefix defined for the VODataService schema; however, instance documents may use any prefix. In applications where common use of prefixes is recommended (such as with the Registry Interface specification [RI]), use of the vs prefix is recommended. Note also that in this document, the `vr` prefix is used to label, as shorthand, a type or element name that is defined in the VOResource schema, as in `vr:Resource`.

> **Note:**
> One reason one may *not* be able to use `vs` to represent the VODataService schema, version 1.1, is because it is already defined to represent VODataService v1.0 and cannot be overridden. At this writing, there are no IVOA applications in which this is the case. Consult [Appendix B](#) for more details on compatibility issues.

As recommend by the VOResource standard [[VOR](#)], the VODataService schema sets `elementFormDefault="unqualified"`. This means that it is not necessary to qualify element names defined in this schema with a namespace prefix (as there are no global elements defined). The only place it is usually needed is as a qualifier to a VODataService type name given as the value of an `xsi:type` attribute.

## 2.2. Summary of Metadata Concepts

The VODataService extension defines four new types of resources. Two inherit directly from `vr:Resource`:

`vs:DataCollection`
This resource declares the existence of a collection of data, what it represents, and how to get it. The access to the data may be limited to a human-readable web page (given by `content/referenceURL`); however, if the contents of the collection are available statically via a URL (e.g. an FTP URL to a directory containing all the files), that URL can be provided. It can also provide pointers to other IVOA registered services that can be used to access the data.

`vs:StandardSTC`
This resource type declares one or more coordinate systems described using STC [[STC](#)] such that each can be assigned a globally unique identifier (based on the IVOA identifier for the resource record itself). This identifier can then be referenced in any other STC description in lieu of a fully described coordinate system. Coordinate systems described in this way become reusable standards once they are registered in an IVOA registry.

The other two resource types represent specialized services:

`vs:DataService`
Inheriting from `vr:Service`, this type is for services that access astronomical data. It adds the ability to describe the data's [coverage](#) of the sky, frequency, and time.

`vs:CatalogService`
Inheriting from `vs:DataService`, this type specifically refers to a service that accesses tabular data. In addition to the coverage information, this type adds the ability to describe the tables and their columns. This is intended for describing services that support the "simple" IVOA data access layer protocols such as Simple Image Access [[SIA](#)] and Simple Cone Search [[SCS](#)].

In general, `coverage` refers to the extent that data samples the measurement range of the sky (space), frequency, and time. The coverage metadata (encoded via the `vs:Coverage` type) has two parts. The first part allows a full STC profile description (via the imported STC element, `<stc:STCResourceProfile>`). The second part captures key coverage metadata defined in the IVOA Resource Metadata standard [RM]. The RM-derived coverage elements can be considered summarizing metadata for many of the details that *may* appear within the STC description, and enables simpler searching of high-level coverage information.

The detailed STC profile contained within the `<stc:STCResourceProfile>` element is capable of describing coverage not only in space, time, and frequency but also velocity and redshift. The profile contains up to three types of component descriptions ([STC], section 4.1): coordinate systems, coordinate values, and coordinate areas or ranges. The first component describes the coordinate systems to which coordinate values, areas, and regions are referenced. While any arbitrary system can be described in this first part, it is expected that most VODataService instances will provide a simple pointer to a predefined system in a registered `vs:StandardSTC` record (using the mechanism summarized in section 3.1.2 below). The coordinate values part will usually be used to describe the coordinate resolution, errors, or typical sizes. The coordinate areas part describes actual regions or ranges covered by the resource in any of the given coordinate systems.

Table descriptions appear within a single `<tableset>` element. This element can in turn can contain one or more `<schema>` element in which each **schema** represents a set of logically related tables. It is not required that that the schema grouping match the underlying database's *catalogs* or *schemas* (as defined in [SQLGuide]), though it may. In some cases, such as when describing the table that is returned from an SIA service, the terms *catalog* and *schema* may have little relevance; in this case, the table can be considered part of a sole "default" schema.

For each table in a schema, one can describe each of the columns, providing such information as its name, type, UCD [UCD], units, and a textual description. Providing this information makes it possible to select a resource based on the kind data contained in its tables.

Finally, the VODataService defines specialized interface type (inheriting from `vr:Interface`) called `vs:ParamHTTP`. This type is used to describe the commonly used interface that is invoked over HTTP as either a GET or a POST [HTTP] in which the arguments are encoded as *name=value* pairs. In addition to the access URL, it can include not only the mime-type of the returned response, it can also enumerate the input arguments that are supported by the service implementation. Much like table columns, one can indicate for each argument the name, the UCD, the data type, the units, whether it is required, and a textual description of the argument. Note that this does not capture any interdependencies between arguments. For example, it cannot indicate if one argument only makes sense in the presence of another argument.

# 3. The VODataService Metadata

This section enumerates the types and elements defined in the VODataService extension schema and describes their meaning. Where a term matches a term in the RM, its meaning is given in terms of the RM definition.

## 3.1. Resource Type Extensions

### 3.1.1. DataCollection

A **data collection**, which is describable with the `vs:DataCollection` resource type, is a logical group of data comprising one or more accessible datasets. A collection can contain any combination of images, spectra, catalogs, time-series, or other data. (In contrast, we talk about a *dataset* as being a set of digitally-encoded data that is normally accessible as a single unit--e.g., a file.)

The `vs:DataCollection` type adds seven additional metadata elements beyond the core VOResource metadata [VOR].

---

**vs:DataCollection Type Schema Definition**

```
<xs:complexType name="DataCollection">
    <xs:complexContent>
        <xs:extension base="vr:Resource">
            <xs:sequence>

                <xs:element name="facility" type="vr:ResourceName"
                            minOccurs="0" maxOccurs="unbounded"/>
                <xs:element name="instrument" type="vr:ResourceName"
                            minOccurs="0" maxOccurs="unbounded"/>
                <xs:element name="rights" type="vr:Rights"
                            minOccurs="0" maxOccurs="unbounded"/>
                <xs:element name="format" type="vs:Format"
                            minOccurs="0" maxOccurs="unbounded"/>
                <xs:element name="coverage" type="vs:Coverage" minOccurs="0"/>
                <xs:element name="tableset" type="vs:TableSet" minOccurs="0">
                    <xs:unique name="DataCollection-schemaName">
                        <xs:selector xpath="schema" />
                        <xs:field xpath="name" />
                    </xs:unique>
                    <xs:unique name="DataCollection-tableName">
                        <xs:selector xpath="schema/table" />
                        <xs:field xpath="name" />
                    </xs:unique>
                <xs:element>
                <xs:element name="accessURL" type="vr:AccessURL" minOccurs="0"/>

            </xs:sequence>
        </xs:extension>
    </xs:complexContent>
</xs:complexType>
```

---

The definition of `<tableset>` element places forces certain names within its description to be unique; these constraints are explained further in 3.3.1.

All of the child elements except `<tableset>` derive from RM terms. Four of the elements--`<facility>`, `<instrument>`, `<rights>`, and `<accessURL>`--are reuses of elements defined in the core VOResource schema, sharing the same syntax and similar semantics. In particular, the meanings of `<facility>` and `<instrument>` in the context of `vs:DataCollection` are different from that in `vr:Organisation` only in that in the former type, they refer to the origin of the data.

| vs:DataCollection Extension Metadata Elements | |
|---|---|
| **Element** | **Definition** |
| facility | *RM Name:* Facility |
| | *Value type:* string with optional ID attribute: `vr:ResourceName` |
| | *Semantic Meaning:* the observatory or facility used to collect the data contained or managed by this resource. |

| vs:DataCollection Extension Metadata Elements | |
|---|---|
| **Element** | **Definition** |
| | Occurrences: optional; multiple occurrences allowed |
| instrument | RM Name: Instrument |
| | Value type: string with optional ID attribute: `vr:ResourceName` |
| | Semantic Meaning: the instrument used to collect the data contained or managed by this resource. |
| | Occurrences: optional; multiple occurrences allowed |
| rights | RM Name: Rights |
| | Value type: string, controlled vocabulary: `xs:token` |
| | Semantic Meaning: Information about rights held in and over the resource. |
| | Occurrences: optional; multiple occurrences allowed |
| | Allowed Values: `public` unrestricted, public access is allowed without authentication. |
| | `secure` authenticated, public access is allowed. |
| | `proprietary` only proprietary access is allowed with authentication. |
| format | RM Name: Format |
| | Value type: string with optional isMIMEType attribute, `vs:Format` |
| | Semantic Meaning: The physical or digital manifestation of the information supported by a resource. |
| | Occurrences: optional; multiple occurrences allowed |
| | Comments: MIME types should be used for network-retrievable, digital data, and the `isMIMEType` attribute should be set to explicitly to "true". Non-MIME type values are used for media that cannot be retrieved over the network--e.g. CDROM, poster, slides, video cassette, etc. |
| coverage | RM Name: Coverage |
| | Value type: composite; `vs:Coverage` |
| | Semantic Meaning: Extent of the content of the resource over space, time, and frequency. |
| | Occurrences: optional |
| tableset | Value type: composite; `vs:TableSet` |
| | Semantic Meaning: A description of tables that are part of this collection. |
| | Occurrences: optional. |
| accessURL | RM Name: Service.AccessURL |
| | Value type: URL with optional `use` attribute: `vr:AccessURL` |
| | Semantic Meaning: The URL can be used to download the data contained in this data collection. |
| | Occurrences: required; multiple occurrences allowed. |

The `vs:Format` type is used for providing a value to the `<format>` element:

---

**vs:Format Type Schema Definition**

```
<xs:complexType name="Format">
    <xs:simpleContent>
        <xs:extension base="xs:token">
            <xs:attribute name="isMIMEType" type="xs:boolean" default="false"/>
        </xs:extension>
    </xs:simpleContent>
</xs:complexType>
```

---

The `isMIMEType` attribute provides a flag to indicate if the value represents an actual MIME-type: if it is, this attribute should be explicitly set to "true".

See section 3.3 for a specification of the `vs:TableSet` type for describing tables.

### 3.1.2. StandardSTC

The `vs:StandardSTC` resource type is used to register standard coordinate systems, positions, or regions using the Space-Time Coordinate (STC, [STC]) standard schema so that they can by uniquely referenced by name by other resource descriptions or applications. This resource type extends the core metadata with a single element, `<stcDefinitions>`, which contains the STC definitions.

---

**vs:StandardSTC Type Schema Definition**

```
<xs:complexType name="StandardSTC">
    <xs:complexContent>
        <xs:extension base="vr:Resource">
            <xs:sequence>
                <xs:element name="stcDefinitions"
                            type="stc:STCResourceProfile"
                            minOccurs="0" maxOccurs="unbounded"/>
            </xs:sequence>
        </xs:extension>
    </xs:complexContent>
</xs:complexType>
```

---

The curation metadata that is part of the core VODataService should generally refer to the publishing organization and persons that are responsible for defining the systems, updating the definitions as needed, and responding to user questions about the definitions. The content metadata, in particular the textual contents of the `<description>` element, should describe the purpose of the definition and where references to the defined systems, positions, or regions may be used.

---

**vs:StandardSTC Extension Metadata Elements**

| Element | Definition | |
|---|---|---|
| stcDefintions | *Value type:* | composite; `stc:stcDescriptionType` |
| | *Semantic Meaning:* | the definitions of systems, positions, and regions that are available for referencing. |
| | *Occurrences:* | required; multiple occurrences allowed |

---

The content of the `<stcDefinitions>` element is controlled by the STC schema. Because that schema uses the `elementFormDefault="true"` and most of the STC elements are defined to be global [schema], `<stcDefinitions>` child elements must be qualified as being in the STC namespace (http://www.ivoa.net/xml/STC/stc-v1.30.xsd), by either setting the default namespace (via the `xmlns` attribute) or via explicit qualification via a prefix (see example).

### 3.1.3. DataService

The `vs:DataService` resource type is for describing a service that provides access to astronomical data. This service adds to the core VOResource service metadata the ability to associate an observing facility and/or instrument with the data as well as describe the coordinate coverage of data via its child `<coverage>` element. Note that while these elements are all optional, a resource of this type still implies access to astronomical data.

| vs:DataService Type Schema Definition |
|---|
| ```
<xs:complexType name="DataService">
    <xs:complexContent>
        <xs:extension base="vr:Service">
            <xs:sequence>
                <xs:element name="facility" type="vr:ResourceName"
                            minOccurs="0" maxOccurs="unbounded"/>
                <xs:element name="instrument" type="vr:ResourceName"
                            minOccurs="0" maxOccurs="unbounded"/>
                <xs:element name="coverage" type="vs:Coverage" minOccurs="0"/>
            </xs:sequence>
        </xs:extension>
    </xs:complexContent>
</xs:complexType>
``` |

The use and meaning of the `<facility>` and `<instrument>` elements are the same as for vs:DataCollection.

| vs:DataService Extension Metadata Elements | | |
|---|---|---|
| **Element** | **Definition** | |
| facility | *RM Name:* | Facility |
| | *Value type:* | string with optional ID attribute: vr:ResourceName |
| | *Semantic Meaning:* | the observatory or facility used to collect the data contained or managed by this resource. |
| | *Occurrences:* | optional; multiple occurrences allowed |
| instrument | *RM Name:* | Instrument |
| | *Value type:* | string with optional ID attribute: vr:ResourceName |
| | *Semantic Meaning:* | the instrument used to collect the data contained or managed by this resource. |
| | *Occurrences:* | optional; multiple occurrences allowed |
| coverage | *RM Name:* | Coverage |
| | *Value type:* | composite; `vs:Coverage` |
| | *Semantic Meaning:* | Extent of the content of the resource over space, time, and frequency. |
| | *Occurrences:* | optional |

The contents of the `<coverage>` element are detailed in section 3.2.

### 3.1.4. CatalogService





The `vs:CatalogService` resource type is for describing a service that interacts with astronomical data through one or more specified tables. Because it extends the `vs:DataService` type, a catalog service can have a coverage description as well. The tabular data may optionally be described via a `<tableset>` element.

---

**vs:CatalogService Type Schema Definition**

```
<xs:complexType name="CatalogService">
    <xs:complexContent>
        <xs:extension base="vs:DataService">
            <xs:sequence>

                <xs:element name="tableset" type="vs:TableSet" minOccurs="0">
                    <xs:unique name="CatalogService-schemaName">
                        <xs:selector xpath="schema" />
                        <xs:field xpath="name" />
                    </xs:unique>
                    <xs:unique name="CatalogService-tableName">
                        <xs:selector xpath="schema/table" />
                        <xs:field xpath="name" />
                    </xs:unique>
                <xs:element>

            </xs:sequence>
        </xs:extension>
    </xs:complexContent>
</xs:complexType>
```

---

The definition of `<tableset>` element forces certain names within its description to be unique; these constraints are explained further in 3.3.1.

**vs:CatalogService Extension Metadata Elements**

| Element | Definition | |
|---------|------------|---|
| tableset | *Value type:* | composite; `vs:TableSet` |
| | *Semantic Meaning:* | A description of the tables that are accessible through this service. |
| | *Occurrences:* | optional |

## 3.2. Coverage

The `vs:Coverage` type describes how the data samples the sky, frequency, and time.

---

**vs:Coverage Type Schema Definition**

```
<xs:complexType name="Coverage">
    <xs:sequence>

        <xs:element ref="stc:STCResourceProfile" minOccurs="0"/>
        <xs:element name="footprint" type="vs:ServiceReference"
                    minOccurs="0"/>
        <xs:element name="waveband" type="vs:Waveband"
                    minOccurs="0" maxOccurs="unbounded"/>
        <xs:element name="regionOfRegard" type="xs:float"
                    minOccurs="0" maxOccurs="unbounded"/>

    </xs:sequence>
</xs:complexType>
```

---





A detailed, systematic description of coverage is provided via the child `<stc:STCResourceProfile>` element, taken from the STC schema, version 1.3, with the namespace, `http://www.ivoa.net/xml/STC/stc-v1.30.xsd` (hereafter referred using the `stc:` prefix). This element is defined in the STC schema as a global element; furthermore, the STC schema sets its global `elementFormDefault="qualified"`. Consequently, the `<stc:STCResourceProfile>` element and all its child elements must be qualified as part of the STC namespace as required by XML Schema [schema]. In applications where common use of XML prefixes is required or encouraged (e.g. the IVOA Registry Interfaces [RI]), use of the `stc:` prefix to represent the STC namespace is encouraged.

> **Note:**
> The STC scheme provides rich mark-up for expressing the details of the coverage. In particular, the mark-up is quite flexible in the units that can be used. For example, spectral coverage can be given in terms of frequency, wavelength, or energy. While it is recommended that the overall description given in the `<stc:STCResourceProfile>` be fairly general and approximate, leveraging the richness for a detailed description is allowed.

The remaining elements provide some summary information about the coverage.

| **vs:Coverage Metadata Elements** | |
|---|---|
| **Element** | **Definition** |
| STCResourceProfile | *Value type:* composite: an `stc:STCResourceProfile` element from the STC schema. |
| | *Semantic Meaning:* The STC description of the location of the resource's data (or behavior on data) on the sky, in time, and in frequency space, including resolution. |
| | *Occurrences:* optional |
| | *Comments:* In general, this description should be approximate; a more precise description can be provided by the service referred to by the `<footprint>` element. |
| footprint | *Value type:* a URL with an optional IVOA identifier attribute: |
| | *Semantic Meaning:* a reference to a footprint service for retrieving precise and up-to-date description of coverage. |
| | *Occurrences:* optional |
| | *Comments:* the `ivo-id` attribute refers to a Service record having a footprint service capability. That is, the record will have a capability element describing the footprint service (see "Note on Footprint Service" below for further discussion). The resource referred to may be the current one. |
| waveband | *RM Name:* Coverage.Spectral |
| | *Value type:* string with controlled vocabulary: `vs:Waveband` |





## vs:Coverage Metadata Elements

| Element | Definition | |
|---|---|---|
| | *Semantic Meaning:* | a named spectral region of the electro-magnetic spectrum that the resource's spectral coverage overlaps with. |
| | *Occurrences:* | optional; multiple occurrences allowed |
| | *Allowed Values:* | |
| | `Radio` | any wavelength > 10 mm (or frequency < 30 GHz) |
| | `Millimeter` | 0.1 mm <= wavelength <= 10 mm; 3000 GHz >= frequency >= 30 GHz. |
| | `Infrared` | 1 micron <= wavelength <= 100 microns |
| | `Optical` | 0.3 microns <= wavelength <= 1 micron; 300 nm <= wavelength <= 1000 nm; 3000 Angstroms <= wavelength <= 10000 Angstroms |
| | `UV` | 0.1 micron <= wavelength <= 0.3 microns; 100 nm <= wavelength <= 300 nm; 1000 Angstroms <= wavelength <= 3000 Angstroms |
| | `EUV` | 100 Angstroms <= wavelength <= 1000 Angstroms; 12 eV <= energy <= 120 eV |
| | `X-ray` | 0.1 Angstroms <= wavelength <= 100 Angstroms; 0.12 keV <= energy <= 120 keV |
| | `Gamma-ray` | energy >= 120 keV |
| regionOfRegard | *RM Name:* | Coverage.RegionOfRegard |
| | *Value type:* | a floating point number: `xs:float` |
| | *Semantic Meaning:* | a single numeric value representing the angle, given in decimal degrees, by which a positional query against this resource should be "blurred" in order to get an appropriate match. |
| | *Occurrences:* | optional |
| | *Comments:* | In the case of image repositories, this value might refer to a typical field-of-view size, or the primary beam size for radio aperture synthesis data. In the case of object catalogs, region of regard should normally be the largest of the typical size of the objects, the astrometric errors in the positions, or the resolution of the data. |

**Note on Footprint Service:**

The `<footprint>` element has been defined in anticipation of a future standard IVOA footprint service protocol that can be used to respond to





> detailed spatial overlap queries. Consequently, in the future, applications may be able to assume the protocol that footprint service URL supports. When an application is unable to make any assumptions, the IVOA Identifier given by the attribute should be resolved and the returned resource description should be searched for a recognized footprint service capability.

## 3.3. Tabular Data

The `vs:TableSet` type can be used to describe a set of tables that are part of a single resource and can be consider functionally all located at a single site.

---

**vs:TableSet Type Schema Definition**

```
<xs:complexType name="TableSet">
    <xs:sequence>
        <xs:element name="schema" type="vs:TableSchema"
                    minOccurs="1" maxOccurs="unbounded"/>
    </xs:sequence>

    <xs:anyAttribute namespace="##other" />
</xs:complexType>
```

---

**vs:TableSet Metadata Elements**

| Element | Definition | |
|---------|------------|--|
| schema | *Value type:* | composite; `vs:TableSchema` |
| | *Semantic Meaning:* | A named description of a set of logically related tables. |
| | *Occurrences:* | required; multiple occurrences are allowed. |
| | *Comments:* | See section 3.3.1 regarding unique names for schemas. |

---

The `vs:TableSchema` type collects tables together that are logically related. For example, a single resource may provide access several major astronomical catalogs (e.g. SDSS, 2MASS, and FIRST) from one site, enabling high-performance cross-correlations between them. Each catalog can be described in a separate `<schema>` element, using the elements from the `vs:TableSchema` type.

---

**vs:TableSchema Type Schema Definition**

```
<xs:complexType name="TableSchema">
    <xs:sequence>

        <xs:element name="name" type="xs:token" minOccurs="1" maxOccurs="1"/>
        <xs:element name="title" type="xs:token" minOccurs="0"/>
        <xs:element name="description" type="xs:token"
                    minOccurs="0" maxOccurs="1"/>
        <xs:element name="utype" type="xs:token" minOccurs="0"/>
        <xs:element name="table" type="vs:Table"
                    minOccurs="0" maxOccurs="unbounded"/>

    </xs:sequence>

    <xs:anyAttribute namespace="##other" />
</xs:complexType>
```

---





| vs:TableSchema Metadata Elements | |
|---|---|
| **Element** | **Definition** |
| name | *Value type:*    string: `xs:token` |
| | *Semantic Meaning:*    A name for the set of tables. |
| | *Occurrences:*    required |
| | *Comments:*    If there is no appropriate logical name associated with this set, the name should be explicitly set to "default". See section 3.3.1 regarding the uniqueness of this name. |
| title | *Value type:*    string: `xs:token` |
| | *Semantic Meaning:*    a descriptive, human-interpretable name for the table set. |
| | *Occurrences:*    optional |
| | *Comments:*    This is used for display purposes and is useful when there are multiple schemas in the context (e.g. within a tableset; otherwise, the resource title could be used instead). Note, however, that there is no requirement regarding uniqueness. If a title is not provided, the schema name can be used for display purposes. |
| description | *Value type:*    string: `xs:token` |
| | *Semantic Meaning:*    A free text description of the tableset that should explain in general how all of the tables are related. |
| | *Occurrences:*    optional |
| utype | *Value type:*    string: `xs:token` |
| | *Semantic Meaning:*    an identifier for a concept in a data model that the data in this schema as a whole represent. |
| | *Occurrences:*    optional |
| | *Comments:*    The format defined in the VOTable standard, section 4.1 [VOTable] is strongly recommended; see "Note on UType Format" below. |
| table | *Value type:*    composite: `vs:Table` |
| | *Semantic Meaning:*    A marked description of one of the tables that makes up the set. |
| | *Occurrences:*    optional; multiple occurrences are allowed. |
| | *Comments:*    See section 3.3.1 regarding unique names for schemas. |

> **Note on UType Format:**
> As of this writing, an IVOA standard for the format of utypes is still under development. As a result, the most definitive documentation of the format is in section 4.1 of the VOTable specification [VOTable], which is expected to be a more general form to be spelled out in the eventual utype standard. Use of that latter standard is recommended once it becomes available.

Each table in a schema is described in detail using the `vs:Table` type.





**vs:TableSchema Type Schema Definition**

```
<xs:complexType name="Table">
    <xs:sequence>

        <xs:element name="name" type="xs:token"
                    minOccurs="1" maxOccurs="1"/>
        <xs:element name="title" type="xs:token" minOccurs="0"/>
        <xs:element name="description" type="xs:token" minOccurs="0"/>
        <xs:element name="utype" type="xs:token" minOccurs="0"/>
        <xs:element name="column" type="vs:TableParam"
                    minOccurs="0" maxOccurs="unbounded"/>

    </xs:sequence>

    <xs:attribute name="type" type="xs:string"/>
    <xs:anyAttribute namespace="##other" />
</xs:complexType>
```

**vs:Table Metadata Elements**

| Element | Definition | |
|---|---|---|
| name | *Value type:* | string: **xs:token** |
| | *Semantic Meaning:* | A fully qualified name for the table. This name should include all catalog or schema prefixes needed to sufficiently uniquely distinguish it in a query to the table. |
| | *Occurrences:* | required |
| | *Comments:* | In general, the format of the qualified name may depend on the on the context; however, when the table is intended to be queryable via ADQL [ADQL], then the catalog and schema qualifiers are delimited from the table name with dots (.). |
| | | If this table is part of the schema named "default", the schema name does not need to appear in this table name, unless it is required by an associated access service. |
| | | If there is no appropriate logical name associated with this table, the name should be explicitly set to "default". See section 3.3.1 regarding the uniqueness of this name. |
| title | *Value type:* | string: **xs:token** |
| | *Semantic Meaning:* | a descriptive, human-interpretable name for the table set. |
| | *Occurrences:* | optional |
| | *Comments:* | This is used for display purposes. There is no requirement regarding uniqueness. If a title is not provided, the table name can be used for display purposes. |
| description | *Value type:* | string: **xs:token** |
| | *Semantic Meaning:* | A free-text description of the table's contents. |





| vs:Table Metadata Elements | | |
|---|---|---|
| **Element** | **Definition** | |
| | *Occurrences:* | optional |
| utype | *Value type:* | string: `xs:token` |
| | *Semantic Meaning:* | an identifier for a concept in a data model that the data in this table as a whole represent. |
| | *Occurrences:* | optional |
| | *Comments:* | The format defined in the VOTable standard, section 4.1 [VOTable] is strongly recommended; see "Note on UType Format" above. |
| column | *Value type:* | composite: `vs:TableParam` |
| | *Semantic Meaning:* | A marked description of one of the table's columns. |
| | *Occurrences:* | optional; multiple occurrences are allowed. |
| | *Comments:* | See section 3.5 for the description of this element's contents. |
| foreignKey | *Value type:* | composite: `vs:ForeignKey` |
| | *Semantic Meaning:* | A description of a foreign keys, one or more columns from the current table that can be used to join with another table. |
| | *Occurrences:* | optional; multiple occurrences are allowed. |
| | *Comments:* | See section 3.5.2 for the description of this element's contents. |

Each column in a table can be described using the `vs:TableParam` type which is described in section 3.5. The foreign keys in the table that can be used to join it with another table can be described with the `vs:ForeignKey` type (section 3.3.2). A foreign key description should only refer to tables described within the current table set.

The `vs:Table` also provides an attribute for indicating the role a table plays in the schema:

| vs:Table Attributes | | |
|---|---|---|
| **Attribute** | **Definition** | |
| `type` | *Value type:* | string: `xs:token` |
| | *Semantic Meaning:* | a name indicating the role this table plays. |
| | *Occurrences:* | optional |
| | *Recommeded Values:* `output` | this table structure is used to format the output from a query |
| | `base_table` | this table contains records that represent the main subjects of the parent schema; other tables contain ancillary data. |
| | `view` | the table represents a useful combination or subset of other tables. |
| | Other values are allowed. | |

### 3.3.1. Unique Names for Tables





The definitions of the `<tableset>` elements used in the `vs:DataCollection` and `vs:CatalogService` types constrain certain names to be unique. In particular, all schema names within a `<tableset>` element must be unique, and all table names within a `<tableset>` element must be unique. (A schema and table may share a common name, such as "default".) These constraints makes it possible to uniquely locate the description of a schema or table within a VOResource description.

---

**Example**

The uniqueness constraints for names within table sets guarantee that when the following XPath queries are applied to a `<tableset>` element, zero or one node only will be returned:

```
schema[@name="default"]
schema/table[@name="default"]
```

---

Name uniqueness is only required when the table set description is part of a VOResource description. The name uniqueness rules *should* also be applied to other uses of the `vs:TableSet` element outside of a VOResource description.

### 3.3.2. Foreign Keys

The `vs:ForeignKey` type allows one to describe foreign keys in a table that allow it to be joined effectively with another table. A foreign key is a set of columns that map to a corresponding set of columns in another table.

---

**vs:ForeignKey Type Schema Definition**

```
<xs:complexType name="ForeignKey">
    <xs:sequence>
        <xs:element name="targetTable" type="xs:token"/>
        <xs:element name="fkColumn" type="vs:FKColumn"
                    minOccurs="1" maxOccurs="unbounded"/>
        <xs:element name="description" type="xs:token" minOccurs="0"/>
        <xs:element name="utype" type="xs:token" minOccurs="0"/>
    </xs:sequence>
</xs:complexType>
```

---

In this model, the source of the foreign key is the current table being described (i.e. represented by the `<table>` element that contains the `vs:ForeignKey` description, and thus doesn't need to be named explicitly). The key that is described points to the table given by the `<targetTable>` child element. Each child `<fkColumn>` element then gives a pair of columns, one from the source table and one from the target table, that can be constrained to be equal in a query that joins the two tables.

**vs:ForeignKey Metadata Elements**

| Element | Definition | |
|---|---|---|
| targetTable | *Value type:* | string: `xs:token` |
| | *Semantic Meaning:* | the fully-qualified name (including catalog and schema, as applicable) of the table that can be joined with the table containing this foreign key. |
| | *Occurrences:* | required |
| fkColumn | *Value type:* | composite: `vs:FKColumn` |





**vs:ForeignKey Metadata Elements**

| Element | Definition | |
|---|---|---|
| | *Semantic Meaning:* | a pair of column names, one from this table and one from the target table that should be used to join the tables in a query. |
| | *Occurrences:* | required; multiple occurrences are allowed. |
| | *Comments:* | There should be one `<fkColumn>` element for each column that makes up the foreign key. |
| description | *Value type:* | string: `xs:token` |
| | *Semantic Meaning:* | a free-text description of what this key points to and what the relationship means |
| | *Occurrences:* | optional |
| utype | *Value type:* | string: `xs:token` |
| | *Semantic Meaning:* | an identifier for a concept in a data model that the association enabled by this key represents. |
| | *Occurrences:* | optional |
| | *Comments:* | The format defined in the VOTable standard, section 4.1 [VOTable] is strongly recommended; see "Note on UType Format" above. |

**vs:FKColumn Type Schema Definition**

```
<xs:complexType name="FKColumn">
    <xs:sequence>
        <xs:element name="fromColumn" type="xs:token"/>
        <xs:element name="targetColumn" type="xs:token"/>
    </xs:sequence>
</xs:complexType>
```

**vs:FKColumn Metadata Elements**

| Element | Definition | |
|---|---|---|
| fromColumn | *Value type:* | string: `xs:token` |
| | *Semantic Meaning:* | The unqualified name of the column from the current table. |
| | *Occurrences:* | required |
| targetColumn | *Value type:* | string: `xs:token` |
| | *Semantic Meaning:* | The unqualified name of the column from the target table. |
| | *Occurrences:* | required |

**Example**

a description of a foreign key in an observation table pointing into a filter table.

```
<tableset>
  <schema>
    <name> LSST </name>
    <table>
      <name> LSST.Filters </name>
```





```
                <description> a description of the filters used in observations </description>
                <column>
                    <name>ID</name>
                    ...
                </column>
                ...
            </table>
            <table>
                <name> LSST.Observations </name>
                <description> a listing of the observations made </description>
                <column>
                    <name>filterID</name>
                    <description>
                       the key into the Filter table pointing to the filter used
                       in the observation.
                    </description>
                    ...
                </column>
                ...
                <foreignKey>
                    <targetTable> LSST.Filters </targetTable>
                    <fkColumn>
                        <fromColumn> filterID </fromColumn>
                        <targetColumn> ID </targetColumn>
                    </fkColumn>
                </foreignKey>
            </table>
        </schema>
    </tableset>
```

### 3.3.3. Extending Table Metadata

It is envisioned that it may be useful in the future to provide richer metadata for describing tables within a VOResource description than what are defined in this document. This document recommends the use of the following extension mechanisms when richer descriptions are desired:

1. Use extended types by applying the `xsi:type` attribute to the `<tableset>`, `<schema>`, `<table>`, `<column>` and/or `<dataType>` elements. The values provided in the attributes must refer to an XML type legally extended from the types associated with these elements according to the rules of XML Schema [schema] and the VOResource specification [VOR].

2. Apply a globally-defined attribute from a schema other than VODataService (i.e. from a namespace other than "http://www.ivoa.net/xml/VODataService/v1.1") to any of the `<tableset>`, `<schema>`, `<table>`, and/or `<column>` elements.

3. When the extended metadata is specific to how the table data is accessed via a particular service protocol, then the new metadata can be incorporated into a specific *capability extension* (as described in the VOResource specification [VOR]). This extension may make use of the various names within the `<tableset>` to indicate where the extension metadata apply.

4. Use the `extendedType` attribute of the `<dataType>` element (see section 3.5.3) to indicate a more specific data type then those defined by the `vs:TableParam` type.

## 3.4. Interface Type Extension: ParamHTTP

The `vs:ParamHTTP` type is a specialized service interface description that extends the





VOResource `vr:Interface` type (as recommended by [VOR], section 2.3.2). It describes a service interface that is invoke over HTTP via a GET or a POST [HTTP] in which the inputs are parameters encoded as *name=value* pairs, delimited by ampersands (`&`) and URL-encoded [URI]. When the service is invoked as a GET service, this input list is appended to a base URL (where the result must form a legal URL. Usually, the URL contains a question mark (`?`) setting off a list of URL arguments to the URL:

---

**Example**

A service that takes 3 parameters: `ra, dec, radius`

```
http://data.archive.edu/cgi-bin/search?ra=12.32&dec=-10.3&radius=0.1
```

---

When the service is invoked as a POST, the encoded list of parameters are uploaded to the service as the HTTP Message Body.

---

**Example**

The above GET request example shown as an HTTP POST message.

```
POST /cgi-bin/search
User-Agent: Python script
Content-Type: application/x-www-form-urlencoded
Content-Lenth: 29

ra=12.32&dec=-10.3&radius=0.1
```

---

The `vs:ParamHTTP` type is intended for (but not limited to) use in describing an interface within a VOResource description of a service capability (as described in [VOR], section 2.2.2); that is, it can be invoked via the `xsi:type` attribute on an `<interface>` element.

---

**vs:ParamHTTP Type Schema Definition**

```xml
<xs:complexType name="ParamHTTP">
    <xs:complexContent>
        <xs:extension base="vr:Interface">
            <xs:sequence>

                <xs:element name="queryType" type="vs:HTTPQueryType"
                            minOccurs="0" maxOccurs="2"/>
                <xs:element name="resultType" type="xs:token"
                            minOccurs="0" maxOccurs="1"/>
                <xs:element name="param" type="vs:InputParam" minOccurs="0"
                             maxOccurs="unbounded"/>
                <xs:element name="testQuery" type="xs:string" minOccurs="0"
                            maxOccurs="unbounded"/>

            </xs:sequence>
        </xs:extension>
    </xs:complexContent>
</xs:complexType>
```

---

The extension metadata defined in the schema definition above are all optional. Nevertheless, even when an `<interface>` instance does not include any of these extended child elements, the use of `xsi:type="vs:ParamHTTP"` indicates that the interface accessed via the URL given by the `<accessURL>` element complies with the general parameter-based protocol described in this section.

---

**vs:ParamHTTP Extension Metadata Elements**





| Element | Definition | |
|---------|-----------|---|
| queryType | *Value type:* | string with controlled values: `vs:HTTPQueryType` |
| | *Semantic Meaning:* | The type of HTTP request supported by the interface, either `GET` or `POST`. |
| | *Occurrences:* | optional; 2 occurrences are allowed to indicate that both GET and POST are supported |
| | *Allowed Values:* | `GET` or `POST` |
| resultType | *Value type:* | a string in MIME type format: `xs:token` |
| | *Semantic Meaning:* | The MIME type of a document returned in the HTTP response. |
| | *Occurrences:* | optional |
| param | *Value type:* | composite: `vs:InputParam` |
| | *Semantic Meaning:* | a description of an input parameter that can be provided as a *name=value* argument to the service. |
| | *Occurrences:* | optional; multiple occurrences allowed |
| | *Comments:* | See section 3.5 for the description of this element's contents. |
| testQuery | *Value type:* | a string in MIME type format: `xs:token` |
| | *Semantic Meaning:* | an ampersand-delimited list of arguments that can be used to test this service interface; when provided as the input to this interface, it will produce a legal, non-null response. |
| | *Occurrences:* | optional; multiple occurrences allowed |
| | *Comments:* | When the interface supports GET, then the full query URL is formed by the concatonation of the base URL (given by the accessURL) and the value given by this testQuery element. |

A important intended use of the `vs:ParamHTTP` type is describing the interface of an IVOA standard service protocol of the "simple" variety, such as the Simple Image Access Protocol [SIA]. In particular, it is recommended that specifications that define how a standard service is registered in a registry *require* the use of the `vs:ParamHTTP` interface type when it is applicable.

Normally, a VOResource description indicates its support for a standard protocol with `<capability>` element having a `standardID` attribute set to specific URI representing the standard. The standard will usually spell out the HTTP query type, the returned MIME type, and input parameters required for compliance; therefore, it is not necessary that the `vs:ParamHTTP` description provide any of the optional extended metadata, as they are already implied by the `standardID`. The description need only reflect the optional or locally unique features of the interface. In particular, description may include

- a `<queryType>` element for a type that is not required by the standard (as long as the required query type is supported as well),

- `<param>` elements for any optional parameters or local extended parameters (when allowed by the standard).

Of course, listing required parameters is always allowed, even when describing a standard





interface as long as these are consistent with the service specification and the corresponding `<param>` elements include the attribute `use="required"` (see 3.5.1). The `<param>` elements for custom parameters that are not part of the standard (but are rather local customizations) should include the attribute `std="false"`.

## 3.5. Data Parameters

The VODataService schema provides several element types for describing different kinds of data parameters used in datasets and services, including service input parameters and table columns. The parameter types allow one to fully describe a parameter in terms of metadata that includes name, data type, and meaning.

All the VODataService parameter types derive from a base type called `vs:BaseParam` which defines all the common parameter metadata except the data type.

| vs:BaseParam Type Schema Definition |
|---|
| ```
<xs:complexType name="BaseParam">
    <xs:sequence>

        <xs:element name="name" type="xs:token" minOccurs="0"/>
        <xs:element name="description" type="xs:token" minOccurs="0"/>
        <xs:element name="unit" type="xs:token" minOccurs="0"/>
        <xs:element name="ucd" type="xs:token" minOccurs="0"/>
        <xs:element name="utype" type="xs:token" minOccurs="0"/>

    </xs:sequence>
    <xs:anyAttribute namespace="##other" />
</xs:complexType>
``` |

| vs:BaseParam Metadata Elements | | |
|---|---|---|
| **Element** | **Definition** | |
| name | *Value type:* | string: `xs:token` |
| | *Semantic Meaning:* | The name of the column. |
| | *Occurrences:* | optional |
| description | *Value type:* | string: `xs:token` |
| | *Semantic Meaning:* | a free-text description of the column's contents |
| | *Occurrences:* | optional |
| unit | *Value type:* | string: `xs:token` |
| | *Semantic Meaning:* | the unit associated with all values associated with this parameter or table column. |
| | *Occurrences:* | optional |
| ucd | *Value type:* | string: `xs:token` |
| | *Semantic Meaning:* | the name of a unified content descriptor that describes the scientific content of the parameter. |
| | *Occurrences:* | optional |
| | *Comments:* | There are no requirements for compliance with any particular UCD standard. The format of the UCD can be used to distinguish between UCD1, UCD1+, and SIA-UCD. See [UCD] for the latest IVOA standard set. |

**vs:BaseParam Metadata Elements**

| Element | Definition | |
|---------|-----------|---|
| utype | *Value type:* | string: `xs:token` |
| | *Semantic Meaning:* | an identifier for a concept in a data model that the data in this schema as a whole represent. |
| | *Occurrences:* | optional |
| | *Comments:* | The format defined in the VOTable standard, section 4.1 [VOTable] is strongly recommended; see "Note on UType Format" above. |

Leaving the data type metadatum out of `vs:BaseParam` allows the different kinds of parameters derived from `vs:BaseParam` to restrict the allowed data types to specific sets. The subsections below describe the different data types associated with input parameters (`vs:InputParam`) and table columns (`vs:TableParam`). The XML types associated with their `<dataType>` elements derive from a common parent, `vs:DataType`.

**vs:DataType Type Schema Definition**

```
<xs:complexType name="DataType">
    <xs:simpleContent>
      <xs:extension base="xs:token">
          <xs:attribute name="arraysize" type="vs:ArrayShape" default="1"/>
          <xs:attribute name="delim" type="xs:string" default=" "/>
          <xs:attribute name="extendedType" type="xs:string"/>
          <xs:attribute name="extendedSchema" type="xs:anyURI"/>
          <xs:anyAttribute namespace="##other" />
      </xs:extension>
    </xs:simpleContent>
</xs:complexType>
```

The content of an element of type `vs:DataType` is the name of the data type for the current parameter. When the element is explicitly a `vs:DataType` (as opposed to one of its derived types), there are no restrictions on the names that may be included.

A data type description can be augmented via a common set of `vs:DataType` attributes, defined below. The `arraysize` attribute indicates the parameter is an array of values of the named type. Its value describes the shape of the array, and the `delim` attribute may be used to indicate the delimiter that should appear between elements of an array value. Depending on the application context, these attribute may not be enough to effectively parse the array values, in which case more information must be brought to bear either through assumptions about a particular derived `vs:DataType` or through additional attributes.

More descriptive information about the type can be provided via `extendedType` and `extendedSchema`, which provide an alternate data type name. It's expected that this name will only be understood by a special subset of applications. The name given in the element content, then, represents a more commonly understood "fall-back" type. Arbitrary information can also be provided via any prefix-qualified, globally defined attribute drawn from an XML Schema other than VODataService (by virtue of the `xs:anyAttribute` specification shown above).

**vs:DataType Attributes**

| Attribute | Definition |
|-----------|-----------|

| vs:DataType Attributes | |
|---|---|
| **Attribute** | **Definition** |
| arraysize | *Value type:* the VOTable arraysize format (`vs:ArrayShape`): $LxMxN...$, where each x-delimited positive integer is a length along a dimension of a multi-dimensional array. A single integer indicates a one dimensional array. Instead of an integer, the last length can be set to "*" which indicates a variable length. |
| | *Semantic Meaning:* The attribute's presence indicates that parameter holds an array values; the attribute's value indicates the length of the array along each dimension of the multi-dimensional array. |
| | *Occurrences:* optional |
| | *Default Value:* 1 (i.e. the parameter value is scalar) |
| delim | *Value type:* string: `xs:string` |
| | *Semantic Meaning:* the string that is used to delimit element of an array value when `arraysize` is not "1". |
| | *Occurrences:* optional. |
| | *Comments:* Unless specifically disallowed by the context, applications should allow optional spaces to appear in an actual data value before and after the delimiter (e.g. "1, 5" when delim=","). |
| extendedType | *Value type:* string: `xs:string`. |
| | *Semantic Meaning:* The data value represented by this type can be interpreted as a custom type identified by the value of this attribute. |
| | *Occurrences:* optional |
| | *Comments:* The name implies a particular expected format for the data value that can be parsed into a value in memory. |
| | If an application does not recognize this extendedType, it should attempt to handle value assuming the type given by the element's value. "string" (or its equivalent) is a recommended default type. |
| | This element may make use of the extendedSchema attribute and/or any arbitrary (qualified) attribute to refine the identification of the type. |
| extendedSchema | *Value type:* URI: `xs:anyURI`. |
| | *Semantic Meaning:* An identifier for the schema that the value given by the extended attribute is drawn from. |
| | *Occurrences:* optional |
| | *Comments:* This attribute is normally ignored if the extended element is not present. |

Note that in the derived parameter description types described below, the `<dataType>` element is optional. Its absence from the parameter description does *not* mean that the parameter can support any data type; rather, it means that the data type simply has not been provided (which may limit what an application can do with the parameter). If a parameter can truly support any data type, the `vs:BaseParam` type can be used directly when the context permits.

### 3.5.1. Input Parameters

Actual parameters are normally described with types derived from `vs:BaseParam`. The `vs:InputParam` is intended for describing an input parameter to a service or function. The allowed data type names (given in the metadata table below) do not imply a size or precise format; rather, they are intended to be sufficient for describing an input paramter to a simple REST-like service or a function in a weakly-typed (e.g. scripting) language.

---

**vs:InputParam Type Schema Definition**

```
<xs:complexType name="InputParam">
    <xs:complexContent>
        <xs:extension base="vs:BaseParam">
            <xs:sequence>
                <xs:element name="dataType" type="vs:SimpleDataType"
                            minOccurs="0"/>
            </xs:sequence>

            <xs:attribute name="use" type="vs:ParamUse" default="optional"/>
            <xs:attribute name="std" type="xs:boolean" default="true"/>
        </xs:extension>
    </xs:complexContent>
</xs:complexType>
```

---

By fixing the `<dataType>` child element to that of the `vs:SimpleDataType`, the possible types are restricted to predefined set appropriate for input parameters.

**vs:InputParam Extension Metadata Elements**

| Element | Definition | |
|---|---|---|
| dataType | *Value type:* | string with optional attributes: `vs:SimpleDataType` |
| | *Semantic Meaning:* | a type of data contained in the column. |
| | *Occurrences:* | optional |
| | *Allowed Values:* | The following type names correspond to the same data types defined in the VOTable standard [VOTable]: `boolean`, `char`, `integer`, `real`, and `complex`. The additional type, `string`, is equivalent to `char` when the attribute `arrayshape="*"`. |

The `vs:InputParam` type accepts two attributes that indicate the role that the parameter plays as input to the service or function:

**vs:InputParam Attributes**

| Attribute | Definition | |
|---|---|---|
| use | *Value type:* | string with controlled values: `vs:ParamUse` |
| | *Semantic Meaning:* | An indication of whether this parameter is required to be provided for the application or service to work properly. |
| | *Occurrences:* | optional |

| vs:InputParam Attributes | | |
|---|---|---|
| **Attribute** | **Definition** | |
| | *Allowed Values:* | `required` the parameter must be provided.<br>`optional` the parameter need not be provided (default). |
| `std` | *Value type:* | `true` or `false` (`xs:boolean`) |
| | *Semantic Meaning:* | If true, the meaning and behavior of this parameter is reserved and defined by a standard interface. If false, it represents an implementation-specific parameter that effectively extends the behavior of the service or application. The default is true. |
| | *Occurrences:* | optional |

---

**Example**

A description of an input parameter that might appear inside an `vs:ParamHTTP` interface description. As noted in [section 3.4](#), a `<param>` element uses the `vs:InputParam` type to describe itself.

```
<param use="required">
  <name> radius </name>
  <description>
    search radius; returned objects are restricted to fall
    within this angular distance of the search position.
  </description>
  <ucd> phys.angSize </ucd>
  <dataType> real </dataType>
</param>
```

### 3.5.2. Table Columns

The `vs:TableParam` is also derived from `vs:BaseParam`, and is designed for describing a column of a table.

---

### vs:TableParam Type Schema Definition

```
<xs:complexType name="TableParam">
    <xs:complexContent>
        <xs:extension base="vs:BaseParam">
            <xs:sequence>

                <xs:element name="dataType" type="vs:TableDataType"
                            minOccurs="0"/>
                <xs:element name="flag" type="xs:token"
                            minOccurs="0" maxOccurs="unbounded"/>

            </xs:sequence>

            <xs:attribute name="std" type="xs:boolean"/>
        </xs:extension>
    </xs:complexContent>
</xs:complexType>

<xs:complexType name="TableDataType" abstract="true">
    <xs:simpleContent>
      <xs:extension base="vs:DataType"/>
    </xs:simpleContent>
</xs:complexType>
```

A table column's data type is indicated with the `<dataType>` element with a name drawn from a standard set of names. The `vs:TableParam` type is not restricted to a single standard set, and the VODataService schema defines two standard sets: one corresponding to VOTable data types [VOTable] and one for Table Access Protocol types. Because its XML type, `vs:TableDataType` is abstract, the `<dataType>` element MUST include an `xsi:type` attribute to indicate which standard set of type names is being used.

---

### Example

A declination column called "Dec" is defined to have the VOTable-defined type, double

```
<column>
    <name> Dec </name>
    <description> the J2000 declination of the object </description>
    <ucd> pos.eq.dec </ucd>
    <dataType xsi:type="vs:VOTableType"> double </dataType>
</column>
```

---

### vs:TableParam Extension Metadata Elements

| Element | | Definition |
|---|---|---|
| dataType | Value type: | string with a required `xsi:type` attribute and additional optional attributes: `vs:TableDataType` |
| | Semantic Meaning: | a type of data contained in the column. |
| | Occurrences: | optional |
| | Allowed Values: | The allowed type names are determined by value of the `xsi:type`; see section 3.5.3 below. |
| flag | Value type: | string with optional attributes: `vs:TableDataType` |
| | Semantic Meaning: | a keyword representing traits of the column. |
| | Occurrences: | optional; multiple occurrences allowed |

| vs:TableParam Extension Metadata Elements | |
|---|---|
| **Element** | **Definition** |
| | *Recommended Values:* `indexed` The column has an index on it for faster search against its values. |
| | `primary` The values column in the column represents in total or in part a primary key for its table. |
| | `nullable` the column may contain null or empty values. |
| | Other values are allowed. |

### 3.5.3. Table Column Data Types

The VODataService schema defines two XML types that derive from `vs:TableDataType`: `vs:VOTableType` and `vs:TAPType`.

| Data Types derived from vs:TableDataType | |
|---|---|
| **Value for xsi:type** | **Definition** |
| vs:VOTableType | *Semantic Meaning:* data types that correspond to the parameter and column types defined in the VOTable schema [VOTable]. |
| | *Allowed Values:* `boolean`, `bit`, `unsignedByte`, `short`, `int`, `long`, `char`, `unicodeChar`, `float`, `double`, `floatComplex`, and `doubleComplex`. String values of arbitrary length are represent by a value of `char` with `arraysize="*"` |
| vs:TAPType | *Semantic Meaning:* data types that correspond column types defined in the Table Access Protocol (v1.0) [TAP]. |
| | *Allowed Values:* `BOOLEAN`, `SMALLINT`, `INTEGER`, `BIGINT`, `REAL`, `DOUBLE`, `TIMESTAMP`, `CHAR`, `VARCHAR`, `BINARY`, `VARBINARY`, `POINT`, `REGION`, `CLOB`, and `BLOB`. String values are represented via `VARCHAR`. |

The `vs:TAPType` XML type provides an additional attribute, `size`, corresponding to the "size" column from the TAP_SCHEMA.columns defined by TAP:

| Additional Attribute for the vs:TAPType set of column data types | |
|---|---|
| **Attribute** | **Definition** |
| `size` | *Value type:* a positive integer: `xs:positiveInteger` |
| | *Semantic Meaning:* The length of the variable-length data type. |
| | *Occurrences:* optional |
| | *Comments:* In the context of TAP, this attribute is only meaning when the data type is `CHAR` or `BINARY`; see discussion below. |

| Example |
|---|
| a representation of a string type using the `vs:VOTableType` set of types: |

```
<column>
    <name> id </name>
    <description> the object identifier </description>
    <ucd> meta.id </ucd>
```

```
        <dataType xsi:type="vs:VOTableType" arraysize="*"> char </dataType>
    </column>
```

the same column described using the `vs:TAPType` set of types:

```
<column>
        <name> id </name>
        <description> the object identifier </description>
        <ucd> meta.id </ucd>
        <dataType xsi:type="vs:TAPType"> VARCHAR </dataType>
</column>
```

the same column again described using the `vs:TAPType` set of types, assuming a fixed-length string:

```
<column>
        <name> id </name>
        <description> the object identifier </description>
        <ucd> meta.id </ucd>
        <dataType xsi:type="vs:TAPType" size="8" > CHAR </dataType>
</column>
```

In general, the `vs:TableParam`'S `<dataType>` can support any non-abstract type legally derived from `vs:TableDataType`. However, in the context of a `vs:DataCollection` or `vs:CatalogService` resource description, it is strongly recommended that either `vs:VOTableType` or `vs:TAPType` (or some other IVOA standard type derived from `vs:TableDataType`) be used to ensure maximum interoperability. When the actual column type is not well matched to a type from one of these standard sets, authors are encouraged to use the `extendedType` attribute to refer to a more specific type. Note that the TAP standard [TAP] defines an explicit mapping between TAP_SCHEMA types and VOTable types. Thus, in the context of a `vs:CatalogService` resource description that supports a TAP capability (perhaps in addition to other catalog services like Simple Cone Search [SCS]), use of the `vs:TAPType` data type is preferred.

> **Note:**
>
> The motivation for providing two standard data type sets, `vs:VOTableType` and `vs:TAPType`, is to maximize the ease of generating the table description, particular as part of the VO Standard Interface [VOSI] and for legacy services. The table description for "stand-alone" SIA, SCS, and SSA services can be readily generated using the `vs:VOTableType` data types from these interface's respective metadata queries. Newer services supporting TAP could generate its description using its TAP_SCHEMA queries.
>
> The motivation for specifying a column's data type using the `xsi:type` mechanism is mainly to allow for the possibility that the official TAP data types will evolve. This allows the IVOA to define new data type sets without updating the VODataService standard. Using non-IVOA-standardized data type names is expected to undermine interoperability and so is therefore discouraged.

## Appendix A: The VODataService XML Schema

### The Complete VOResource Schema

```
<?xml version="1.0" encoding="UTF-8"?>
<xs:schema targetNamespace="http://www.ivoa.net/xml/VODataService/v1.1"
          xmlns:xs="http://www.w3.org/2001/XMLSchema"
```

```
                    xmlns:vr="http://www.ivoa.net/xml/VOResource/v1.0"
                    xmlns:vs="http://www.ivoa.net/xml/VODataService/v1.1"
                    xmlns:stc="http://www.ivoa.net/xml/STC/stc-v1.30.xsd"
                    xmlns:vm="http://www.ivoa.net/xml/VOMetadata/v0.1"
                    elementFormDefault="unqualified" attributeFormDefault="unqualified"
                    version="1.1pr2">

    <xs:annotation>
       <xs:appinfo>
          <vm:schemaName>VODataService</vm:schemaName>
          <vm:schemaPrefix>xs</vm:schemaPrefix>
          <vm:targetPrefix>vs</vm:targetPrefix>
       </xs:appinfo>
       <xs:documentation>
          An extension to the core resource metadata (VOResource) for
          describing data collections and services.
          </xs:documentation>
    </xs:annotation>

    <xs:import namespace="http://www.ivoa.net/xml/VOResource/v1.0"
               schemaLocation="http://www.ivoa.net/xml/VOResource/v1.0"/>
    <xs:import namespace="http://www.ivoa.net/xml/STC/stc-v1.30.xsd"
               schemaLocation="http://www.ivoa.net/xml/STC/stc-v1.30.xsd"/>

    <xs:complexType name="DataCollection">
       <xs:annotation>
          <xs:documentation>
             A logical grouping of data which, in general, is composed of one
             or more accessible datasets.  A collection can contain any
             combination of images, spectra, catalogs, or other data.
             </xs:documentation>
          <xs:documentation>
             (A dataset is a collection of digitally-encoded data that
             is normally accessible as a single unit, e.g. a file.)
             </xs:documentation>
       </xs:annotation>

       <xs:complexContent>
          <xs:extension base="vr:Resource">
             <xs:sequence>

                <xs:element name="facility" type="vr:ResourceName"
                            minOccurs="0" maxOccurs="unbounded">
                   <xs:annotation>
                      <xs:appinfo>
                         <vm:dcterm>Subject</vm:dcterm>
                      </xs:appinfo>
                      <xs:documentation>
                         the observatory or facility used to collect the data
                         contained or managed by this resource.
                         </xs:documentation>
                   </xs:annotation>
                </xs:element>

                <xs:element name="instrument" type="vr:ResourceName"
                            minOccurs="0" maxOccurs="unbounded">
                   <xs:annotation>
                      <xs:appinfo>
                         <vm:dcterm>Subject</vm:dcterm>
                         <vm:dcterm>Subject.Instrument</vm:dcterm>
                      </xs:appinfo>
                      <xs:documentation>
                         the Instrument used to collect the data contain or
                         managed by a resource.
                         </xs:documentation>
                   </xs:annotation>
```

```
                        </xs:element>

                        <xs:element name="rights" type="vr:Rights"
                                    minOccurs="0" maxOccurs="unbounded">
                            <xs:annotation>
                                <xs:appinfo>
                                    <vm:dcterm>Rights</vm:dcterm>
                                </xs:appinfo>
                                <xs:documentation>
                                    Information about rights held in and over the resource.
                                </xs:documentation>
                                <xs:documentation>
                                    This should be repeated for all Rights values that apply.
                                </xs:documentation>
                            </xs:annotation>
                        </xs:element>

                        <xs:element name="format" type="vs:Format"
                                    minOccurs="0" maxOccurs="unbounded">
                            <xs:annotation>
                                <xs:documentation>
                                    The physical or digital manifestation of the information
                                    supported by a resource.
                                </xs:documentation>
                                <xs:documentation>
                                    MIME types should be used for network-retrievable, digital
                                    data.  Non-MIME type values are used for media that cannot
                                    be retrieved over the network--e.g. CDROM, poster, slides,
                                    video cassette, etc.
                                </xs:documentation>
                            </xs:annotation>
                        </xs:element>

                        <xs:element name="coverage" type="vs:Coverage" minOccurs="0">
                            <xs:annotation>
                                <xs:documentation>
                                    Extent of the content of the resource over space, time,
                                    and frequency.
                                </xs:documentation>
                            </xs:annotation>
                        </xs:element>
                        <xs:element name="tableset" type="vs:TableSet" minOccurs="0">
                            <xs:annotation>
                                <xs:documentation>
                                    A description of the tables that are part of this
                                    collection.
                                </xs:documentation>
                                <xs:documentation>
                                    Each schema name and each table name must be
                                    unique within this tableset.
                                </xs:documentation>
                            </xs:annotation>

                            <xs:unique name="DataCollection-schemaName">
                                <xs:selector xpath="schema" />
                                <xs:field xpath="name" />
                            </xs:unique>

                            <xs:unique name="DataCollection-tableName">
                                <xs:selector xpath="schema/table" />
                                <xs:field xpath="name" />
                            </xs:unique>
                        </xs:element>

                        <xs:element name="accessURL" type="vr:AccessURL" minOccurs="0">
```

```
                    <xs:annotation>
                      <xs:documentation>
                        The URL that can be used to download the data contained in
                        this data collection.
                      </xs:documentation>
                    </xs:annotation>
                </xs:element>

            </xs:sequence>
          </xs:extension>
        </xs:complexContent>
</xs:complexType>

<xs:complexType name="Coverage">
    <xs:annotation>
        <xs:documentation>
          A description of how a resource's contents or behavior maps
          to the sky, to time, and to frequency space, including
          coverage and resolution.
        </xs:documentation>
    </xs:annotation>

    <xs:sequence>

        <xs:element ref="stc:STCResourceProfile" minOccurs="0">
            <xs:annotation>
                <xs:documentation>
                  The STC description of the location of the resource's
                  data (or behavior on data) on the sky, in time, and in
                  frequency space, including resolution.
                </xs:documentation>
                <xs:documentation>
                  In general, this description should be approximate; a
                  more precise description can be provided by the
                  footprint service.
                </xs:documentation>
            </xs:annotation>
        </xs:element>

        <xs:element name="footprint" type="vs:ServiceReference"
                    minOccurs="0">
            <xs:annotation>
                <xs:documentation>
                  a reference to a footprint service for retrieving
                  precise and up-to-date description of coverage.
                </xs:documentation>
                <xs:documentation>
                  the ivo-id attribute refers to a Service record
                  that describes the Footprint capability.  That is,
                  the record will have a capability element describing
                  the service.  The resource refered to may be the
                  current one.
                </xs:documentation>
            </xs:annotation>
        </xs:element>

        <xs:element name="waveband" type="vs:Waveband"
                    minOccurs="0" maxOccurs="unbounded">
            <xs:annotation>
                <xs:appinfo>
                  <vm:dcterm>Coverage.Spectral</vm:dcterm>
                </xs:appinfo>
                <xs:documentation>
                  a named spectral region of the electro-magnetic spectrum
                  that the resource's spectral coverage overlaps with.
                </xs:documentation>
```

```
                    </xs:annotation>
                </xs:element>

                <xs:element name="regionOfRegard" type="xs:float" minOccurs="0">
                    <xs:annotation>
                        <xs:appinfo>
                            <vm:dcterm>Coverage.RegionOfRegard</vm:dcterm>
                        </xs:appinfo>
                        <xs:documentation>
                            a single numeric value representing the angle, given
                            in decimal degrees, by which a positional query
                            against this resource should be "blurred" in order
                            to get an appropriate match.
                        </xs:documentation>
                        <xs:documentation>
                            In the case of image repositories, it might refer to
                            a typical field-of-view size, or the primary beam
                            size for radio aperture synthesis data.  In the case
                            of object catalogs RoR should normally be the
                            largest of the typical size of the objects, the
                            astrometric errors in the positions, or the
                            resolution of the data.
                        </xs:documentation>
                    </xs:annotation>
                </xs:element>

            </xs:sequence>
        </xs:complexType>

        <xs:complexType name="ServiceReference">
            <xs:annotation>
                <xs:documentation>
                    the service URL for a potentially registerd service.  That is,
                    if an IVOA identifier is also provided, then the service is
                    described in a registry.
                </xs:documentation>
            </xs:annotation>

            <xs:simpleContent>
                <xs:extension base="xs:anyURI">

                    <xs:attribute name="ivo-id" type="vr:IdentifierURI">
                        <xs:annotation>
                            <xs:documentation>
                                The URI form of the IVOA identifier for the service
                                describing the capability refered to by this element.
                            </xs:documentation>
                        </xs:annotation>
                    </xs:attribute>

                </xs:extension>
            </xs:simpleContent>
        </xs:complexType>

        <xs:simpleType name="Waveband">
            <xs:restriction base="xs:token">
                <xs:enumeration value="Radio">
                    <xs:annotation>
                        <xs:documentation>
                            wavelength >= 10 mm; frequency <= 30 GHz.
                        </xs:documentation>
                    </xs:annotation>
                </xs:enumeration>
                <xs:enumeration value="Millimeter">
                    <xs:annotation>
                        <xs:documentation>
```

```
                      0.1 mm <= wavelength <= 10 mm;
                      3000 GHz >= frequency >= 30 GHz.
                    </xs:documentation>
                </xs:annotation>
            </xs:enumeration>
            <xs:enumeration value="Infrared">
                <xs:annotation>
                    <xs:documentation>
                      1 micron <= wavelength <= 100 micons
                    </xs:documentation>
                </xs:annotation>
            </xs:enumeration>
            <xs:enumeration value="Optical">
                <xs:annotation>
                    <xs:documentation>
                      0.3 microns <= wavelength <= 1 micon;
                      300 nm <= wavelength <= 1000 nm;
                      3000 Angstroms <= wavelength <= 10000 Angstroms
                    </xs:documentation>
                </xs:annotation>
            </xs:enumeration>
            <xs:enumeration value="UV">
                <xs:annotation>
                    <xs:documentation>
                      0.1 microns <= wavelength <= 0.3 micon;
                      1000 nm <= wavelength <= 3000 nm;
                      1000 Angstroms <= wavelength <= 30000 Angstroms
                    </xs:documentation>
                </xs:annotation>
            </xs:enumeration>
            <xs:enumeration value="EUV">
                <xs:annotation>
                    <xs:documentation>
                      100 Angstroms <= wavelength <= 1000 Angstroms;
                      12 eV <= energy <= 120 eV
                    </xs:documentation>
                </xs:annotation>
            </xs:enumeration>
            <xs:enumeration value="X-ray">
                <xs:annotation>
                    <xs:documentation>
                      0.1 Angstroms <= wavelength <= 100 Angstroms;
                      0.12 keV <= energy <= 120 keV
                    </xs:documentation>
                </xs:annotation>
            </xs:enumeration>
            <xs:enumeration value="Gamma-ray">
                <xs:annotation>
                    <xs:documentation>
                      energy >= 120 keV
                    </xs:documentation>
                </xs:annotation>
            </xs:enumeration>
        </xs:restriction>
    </xs:simpleType>

    <xs:complexType name="TableSet">
        <xs:annotation>
            <xs:documentation>
              The set of tables hosted by a resource.
            </xs:documentation>
        </xs:annotation>

        <xs:sequence>

            <xs:element name="schema" type="vs:TableSchema"
```

```
                              minOccurs="1" maxOccurs="unbounded">
             <xs:annotation>
               <xs:documentation>
                 A named description of a set of logically related tables.
               </xs:documentation>
               <xs:documentation>
                 The name given by the "name" child element must
                 be unique within this TableSet instance.  If there is
                 only one schema in this set and/or there's no locally
                 appropriate name to provide, the name can be set to
                 "default".
               </xs:documentation>
               <xs:documentation>
                 This aggregation does not need to map to an
                 actual database, catalog, or schema, though the
                 publisher may choose to aggregate along such
                 designations, or particular service protocol may
                 recommend it.
               </xs:documentation>
             </xs:annotation>
         </xs:element>

      </xs:sequence>

   <xs:anyAttribute namespace="##other" />
</xs:complexType>

<xs:complexType name="TableSchema">
   <xs:annotation>
      <xs:documentation>
         A detailed description of a logically-related set of tables
      </xs:documentation>
   </xs:annotation>

   <xs:sequence>
      <xs:element name="name" type="xs:token" minOccurs="1" maxOccurs="1">
         <xs:annotation>
           <xs:documentation>
             A name for the set of tables.
           </xs:documentation>
           <xs:documentation>
             This is used to uniquely identify the table set among
             several table sets.  If a title is not present, this
             name can be used for display purposes.
           </xs:documentation>
           <xs:documentation>
             If there is no appropriate logical name associated with
             this set, the name should be explicitly set to
             "default".
           </xs:documentation>
         </xs:annotation>
      </xs:element>

      <xs:element name="title" type="xs:token" minOccurs="0">
         <xs:annotation>
           <xs:documentation>
             a descriptive, human-interpretable name for the table set.
           </xs:documentation>
           <xs:documentation>
              This is used for display purposes.  There is no requirement
              regarding uniqueness.  It is useful when there are
              multiple schemas in the context (e.g. within a
              tableset; otherwise, the resource title could be
              used instead).
           </xs:documentation>
         </xs:annotation>
```

```
        </xs:element>

        <xs:element name="description" type="xs:token"
                    minOccurs="0" maxOccurs="1">
          <xs:annotation>
            <xs:documentation>
              A free text description of the tableset that should
              explain in general how all of the tables are related.
            </xs:documentation>
          </xs:annotation>
        </xs:element>

        <xs:element name="utype" type="xs:token" minOccurs="0">
          <xs:annotation>
            <xs:documentation>
               an identifier for a concept in a data model that
               the data in this schema as a whole represent.
            </xs:documentation>
            <xs:documentation>
               The format defined in the VOTable standard is strongly
               recommended.
            </xs:documentation>
          </xs:annotation>
        </xs:element>

        <xs:element name="table" type="vs:Table"
                    minOccurs="0" maxOccurs="unbounded">
          <xs:annotation>
            <xs:documentation>
              A description of one of the tables that makes up the set.
            </xs:documentation>
            <xs:documentation>
              The table names for the table should be unique.
            </xs:documentation>
          </xs:annotation>
        </xs:element>

     </xs:sequence>

     <xs:anyAttribute namespace="##other" />
  </xs:complexType>

  <xs:complexType name="Format">
     <xs:simpleContent>
        <xs:extension base="xs:token">
           <xs:attribute name="isMIMEType" type="xs:boolean" default="false">
              <xs:annotation>
                 <xs:documentation>
                    if true, then the content is a MIME Type
                 </xs:documentation>
              </xs:annotation>
           </xs:attribute>
        </xs:extension>
     </xs:simpleContent>
  </xs:complexType>

  <xs:complexType name="DataService">
     <xs:annotation>
        <xs:documentation>
          A service for accessing astronomical data
        </xs:documentation>
     </xs:annotation>

     <xs:complexContent>
        <xs:extension base="vr:Service">
           <xs:sequence>
```

```
            <xs:element name="facility" type="vr:ResourceName"
                        minOccurs="0" maxOccurs="unbounded">
              <xs:annotation>
                <xs:appinfo>
                  <vm:dcterm>Subject</vm:dcterm>
                </xs:appinfo>
                <xs:documentation>
                  the observatory or facility used to collect the data
                  contained or managed by this resource.
                </xs:documentation>
              </xs:annotation>
            </xs:element>

            <xs:element name="instrument" type="vr:ResourceName"
                        minOccurs="0" maxOccurs="unbounded">
              <xs:annotation>
                <xs:appinfo>
                  <vm:dcterm>Subject</vm:dcterm>
                  <vm:dcterm>Subject.Instrument</vm:dcterm>
                </xs:appinfo>
                <xs:documentation>
                  the Instrument used to collect the data contain or
                  managed by a resource.
                </xs:documentation>
              </xs:annotation>
            </xs:element>

            <xs:element name="coverage" type="vs:Coverage" minOccurs="0">
              <xs:annotation>
                <xs:documentation>
                  Extent of the content of the resource over space, time,
                  and frequency.
                </xs:documentation>
              </xs:annotation>
            </xs:element>

          </xs:sequence>
        </xs:extension>
      </xs:complexContent>
</xs:complexType>

<xs:complexType name="ParamHTTP">
    <xs:annotation>
      <xs:documentation>
        A service invoked via an HTTP Query (either Get or Post)
        with a set of arguments consisting of keyword name-value pairs.
      </xs:documentation>
      <xs:documentation>
        Note that the URL for help with this service can be put into
        the Service/ReferenceURL element.
      </xs:documentation>
    </xs:annotation>

    <xs:complexContent>
        <xs:extension base="vr:Interface">
            <xs:sequence>
                <xs:element name="queryType" type="vs:HTTPQueryType"
                            minOccurs="0" maxOccurs="2">
                    <xs:annotation>
                        <xs:documentation>
                          The type of HTTP request, either GET or POST.
                        </xs:documentation>
                        <xs:documentation>
                          The service may indicate support for both GET
                          and POST by providing 2 queryType elements, one
                          with GET and one with POST.
```

```
                    </xs:documentation>
                </xs:annotation>
            </xs:element>

            <xs:element name="resultType" type="xs:token"
                        minOccurs="0" maxOccurs="1">
                <xs:annotation>
                    <xs:documentation>
                      The MIME type of a document returned in the HTTP response.
                    </xs:documentation>
                </xs:annotation>
            </xs:element>

            <xs:element name="param" type="vs:InputParam" minOccurs="0"
                        maxOccurs="unbounded">
                <xs:annotation>
                    <xs:documentation>
                      a description of a input parameter that can be
                      provided as a name=value argument to the service.
                    </xs:documentation>
                </xs:annotation>
            </xs:element>

            <xs:element name="testQuery" type="xs:string" minOccurs="0"
                        maxOccurs="unbounded">
                <xs:annotation>
                    <xs:documentation>
                      a ampersand-delimited list of arguments that
                      can be used to test this service interface;
                      when provided as the input to this interface,
                      it will produce a legal, non-null response.
                    </xs:documentation>
                    <xs:documentation>
                      When then interface supports GET, then the full
                      query URL is formed by the concatonation of the
                      base URL (given by the accessURL) and the value
                      given by this testQuery element.
                    </xs:documentation>
                </xs:annotation>
            </xs:element>

        </xs:sequence>
      </xs:extension>
    </xs:complexContent>
</xs:complexType>

<xs:simpleType name="HTTPQueryType">
    <xs:annotation>
        <xs:documentation>
          The type of HTTP request, either GET or POST.
        </xs:documentation>
    </xs:annotation>
    <xs:restriction base="xs:token">
        <xs:enumeration value="GET"/>
        <xs:enumeration value="POST"/>
    </xs:restriction>
</xs:simpleType>

<xs:complexType name="CatalogService">
    <xs:annotation>
        <xs:documentation>
          A service that interacts with with astronomical data
          through one or more specified tables.
        </xs:documentation>
        <xs:documentation>
          A table with sky coverage typically have columns that give
```

```
                         longitude-latitude positions in some coordinate system.
                     </xs:documentation>
                 </xs:annotation>

             <xs:complexContent>
                 <xs:extension base="vs:DataService">
                     <xs:sequence>
                         <xs:element name="tableset" type="vs:TableSet" minOccurs="0">
                             <xs:annotation>
                                 <xs:documentation>
                                   A description of the tables that are accessible
                                   through this service.
                                 </xs:documentation>
                                 <xs:documentation>
                                   Each schema name and each table name must be
                                   unique within this tableset.
                                 </xs:documentation>
                              </xs:annotation>

                             <xs:unique name="CatalogService-schemaName">
                                 <xs:selector xpath="schema" />
                                 <xs:field xpath="name" />
                             </xs:unique>

                             <xs:unique name="CatalogService-tableName">
                                 <xs:selector xpath="schema/table" />
                                 <xs:field xpath="name" />
                             </xs:unique>
                         </xs:element>
                     </xs:sequence>
                 </xs:extension>
             </xs:complexContent>
         </xs:complexType>

         <xs:complexType name="Table">
             <xs:sequence>
                 <xs:element name="name" type="xs:token"
                             minOccurs="1" maxOccurs="1">
                     <xs:annotation>
                         <xs:documentation>
                            the fully qualified name of the table.  This name
                            should include all catalog or schema prefixes
                            needed to sufficiently uniquely distinguish it in a
                            query.
                         </xs:documentation>
                         <xs:documentation>
                            In general, the format of the qualified name may
                            depend on the context; however, when the
                            table is intended to be queryable via ADQL, then the
                            catalog and schema qualifiers are delimited from the
                            table name with dots (.).
                         </xs:documentation>
                     </xs:annotation>
                 </xs:element>

                 <xs:element name="title" type="xs:token" minOccurs="0">
                     <xs:annotation>
                         <xs:documentation>
                            a descriptive, human-interpretable name for the table.
                         </xs:documentation>
                         <xs:documentation>
                            This is used for display purposes.  There is no requirement
                            regarding uniqueness.
                         </xs:documentation>
                     </xs:annotation>
                 </xs:element>
```

```
            <xs:element name="description" type="xs:token" minOccurs="0">
              <xs:annotation>
                <xs:documentation>
                  a free-text description of the table's contents
                </xs:documentation>
              </xs:annotation>
            </xs:element>

            <xs:element name="utype" type="xs:token" minOccurs="0">
              <xs:annotation>
                <xs:documentation>
                  an identifier for a concept in a data model that
                  the data in this table represent.
                </xs:documentation>
                <xs:documentation>
                  The format defined in the VOTable standard is highly
                  recommended.
                </xs:documentation>
              </xs:annotation>
            </xs:element>

            <xs:element name="column" type="vs:TableParam"
                        minOccurs="0" maxOccurs="unbounded">
              <xs:annotation>
                <xs:documentation>
                  a description of a table column.
                </xs:documentation>
              </xs:annotation>
            </xs:element>

            <xs:element name="foreignKey" type="vs:ForeignKey"
                        minOccurs="0" maxOccurs="unbounded" >
              <xs:annotation>
                <xs:documentation>
                  a description of a foreign keys, one or more columns
                  from the current table that can be used to join with
                  another table.
                </xs:documentation>
              </xs:annotation>
            </xs:element>

          </xs:sequence>

          <xs:attribute name="type" type="xs:string">
            <xs:annotation>
              <xs:documentation>
                a name for the role this table plays.  Recognized
                values include "output", indicating this table is output
                from a query; "base_table", indicating a table
                whose records represent the main subjects of its
                schema; and "view", indicating that the table represents
                a useful combination or subset of other tables.  Other
                values are allowed.
              </xs:documentation>
            </xs:annotation>
          </xs:attribute>

          <xs:anyAttribute namespace="##other" />
      </xs:complexType>

      <xs:complexType name="BaseParam">
        <xs:annotation>
          <xs:documentation>
            a description of a parameter that places no restriction on
            the parameter's data type.
```

```
            </xs:documentation>
            <xs:documentation>
               As the parameter's data type is usually important, schemas
               normally employ a sub-class of this type (e.g. Param),
               rather than this type directly.
            </xs:documentation>
         </xs:annotation>

         <xs:sequence>
            <xs:element name="name" type="xs:token" minOccurs="0">
               <xs:annotation>
                  <xs:documentation>
                     the name of the column
                  </xs:documentation>
               </xs:annotation>
            </xs:element>

            <xs:element name="description" type="xs:token" minOccurs="0">
               <xs:annotation>
                  <xs:documentation>
                     a free-text description of the column's contents
                  </xs:documentation>
               </xs:annotation>
            </xs:element>

            <xs:element name="unit" type="xs:token" minOccurs="0">
               <xs:annotation>
                  <xs:documentation>
                     the unit associated with all values in the column
                  </xs:documentation>
               </xs:annotation>
            </xs:element>

            <xs:element name="ucd" type="xs:token" minOccurs="0">
               <xs:annotation>
                  <xs:documentation>
                     the name of a unified content descriptor that
                     describes the scientific content of the parameter.
                  </xs:documentation>
                  <xs:documentation>
                     There are no requirements for compliance with any
                     particular UCD standard.  The format of the UCD can
                     be used to distinguish between UCD1, UCD1+, and
                     SIA-UCD.  See
                     http://www.ivoa.net/Documents/latest/UCDlist.html
                     for the latest IVOA standard set.
                  </xs:documentation>
               </xs:annotation>
            </xs:element>

            <xs:element name="utype" type="xs:token" minOccurs="0">
               <xs:annotation>
                  <xs:documentation>
                     an identifier for a concept in a data model that
                     the data in this schema represent.
                  </xs:documentation>
                  <xs:documentation>
                     The format defined in the VOTable standard is highly
                     recommended.
                  </xs:documentation>
               </xs:annotation>
            </xs:element>

         </xs:sequence>

         <xs:anyAttribute namespace="##other" />
```

```
          </xs:complexType>

          <xs:complexType name="TableParam">
              <xs:annotation>
                  <xs:documentation>
                      a description of a table parameter having a fixed data type.
                  </xs:documentation>
                  <xs:documentation>
                      The allowed data type names match those supported by VOTable.
                  </xs:documentation>
              </xs:annotation>

              <xs:complexContent>
                  <xs:extension base="vs:BaseParam">
                      <xs:sequence>
                          <xs:element name="dataType" type="vs:TableDataType"
                                      minOccurs="0">
                              <xs:annotation>
                                  <xs:documentation>
                                      a type of data contained in the column
                                  </xs:documentation>
                              </xs:annotation>
                          </xs:element>

                          <xs:element name="flag" type="xs:token"
                                      minOccurs="0" maxOccurs="unbounded">
                              <xs:annotation>
                                  <xs:documentation>
                                      a keyword representing traits of the column.
                                      Recognized values include "indexed", "primary", and
                                      "nullable".
                                  </xs:documentation>
                                  <xs:documentation>
                                      See the specification document for definitions
                                      of recognized keywords.
                                  </xs:documentation>
                              </xs:annotation>
                          </xs:element>
                      </xs:sequence>

                      <xs:attribute name="std" type="xs:boolean">
                          <xs:annotation>
                              <xs:documentation>
                                  If true, the meaning and use of this parameter is
                                  reserved and defined by a standard model.  If false,
                                  it represents a database-specific parameter
                                  that effectively extends beyond the standard.  If
                                  not provided, then the value is unknown.
                              </xs:documentation>
                          </xs:annotation>
                      </xs:attribute>
                  </xs:extension>
              </xs:complexContent>
          </xs:complexType>

          <xs:complexType name="InputParam">
              <xs:annotation>
                  <xs:documentation>
                      a description of a service or function parameter having a
                      fixed data type.
                  </xs:documentation>
                  <xs:documentation>
                      The allowed data type names do not imply a size or precise
                      format.  This type is intended to be sufficient for describing
                      an input parameter to a simple REST service or a function
                      written in a weakly-typed (e.g., scripting) language.
```

```
                    </xs:documentation>
                </xs:annotation>

            <xs:complexContent>
                <xs:extension base="vs:BaseParam">
                    <xs:sequence>
                        <xs:element name="dataType" type="vs:SimpleDataType"
                                    minOccurs="0">
                            <xs:annotation>
                                <xs:documentation>
                                    a type of data contained in the column
                                </xs:documentation>
                            </xs:annotation>
                        </xs:element>
                    </xs:sequence>

                    <xs:attribute name="use" type="vs:ParamUse" default="optional">
                        <xs:annotation>
                            <xs:documentation>
                                An indication of whether this parameter is
                                required to be provided for the application
                                or service to work properly.
                            </xs:documentation>
                            <xs:documentation>
                                Allowed values are "required" and "optional".
                            </xs:documentation>
                        </xs:annotation>
                    </xs:attribute>

                    <xs:attribute name="std" type="xs:boolean" default="true">
                        <xs:annotation>
                            <xs:documentation>
                                If true, the meaning and behavior of this parameter is
                                reserved and defined by a standard interface.  If
                                false, it represents an implementation-specific
                                parameter that effectively extends the behavior of the
                                service or application.
                            </xs:documentation>
                        </xs:annotation>
                    </xs:attribute>

                </xs:extension>
            </xs:complexContent>
        </xs:complexType>

        <xs:simpleType name="ParamUse">
            <xs:restriction base="xs:string">
                <xs:enumeration value="required">
                    <xs:annotation>
                        <xs:documentation>
                            the parameter is required for the application or
                            service to work properly.
                        </xs:documentation>
                    </xs:annotation>
                </xs:enumeration>
                <xs:enumeration value="optional">
                    <xs:annotation>
                        <xs:documentation>
                            the parameter is optional but supported by the application or
                            service.
                        </xs:documentation>
                    </xs:annotation>
                </xs:enumeration>
                <xs:enumeration value="ignored">
                    <xs:annotation>
                        <xs:documentation>
```

```
                    the parameter is not supported and thus is ignored by the
                    application or service.
                </xs:documentation>
            </xs:annotation>
        </xs:enumeration>
    </xs:restriction>
</xs:simpleType>

<xs:complexType name="DataType">
    <xs:annotation>
        <xs:documentation>
            a type (in the computer language sense) associated with a
            parameter with an arbitrary name
        </xs:documentation>
        <xs:documentation>
            This XML type is used as a parent for defining data types
            with a restricted set of names.
        </xs:documentation>
    </xs:annotation>
    <xs:simpleContent>
        <xs:extension base="xs:token">
            <xs:attribute name="arraysize" type="vs:ArrayShape" default="1">
                <xs:annotation>
                    <xs:documentation>
                        the shape of the array that constitutes the value
                    </xs:documentation>
                    <xs:documentation>
                        the default is "1"; i.e. the value is a scalar.
                    </xs:documentation>
                </xs:annotation>
            </xs:attribute>

            <xs:attribute name="delim" type="xs:string" default=" ">
                <xs:annotation>
                    <xs:documentation>
                        the string that is used to delimit elements of an array
                        value when arraysize is not "1".
                    </xs:documentation>
                    <xs:documentation>
                        Unless specifically disallowed by the context,
                        applications should allow optional spaces to
                        appear in an actual data value before and after
                        the delimiter (e.g. "1, 5" when delim=",").
                    </xs:documentation>
                    <xs:documentation>
                        the default is " "; i.e. the values are delimited
                        by spaces.
                    </xs:documentation>
                </xs:annotation>
            </xs:attribute>

            <xs:attribute name="extendedType" type="xs:string">
                <xs:annotation>
                    <xs:documentation>
                        The data value represented by this type can be
                        interpreted as of a custom type identified by
                        the value of this attribute.
                    </xs:documentation>
                    <xs:documentation>
                        If an application does not recognize this
                        extendedType, it should attempt to handle value
                        assuming the type given by the element's value.
                        string is a recommended default type.
                    </xs:documentation>
                    <xs:documentation>
                        This element may make use of the extendedSchema
```

```
                            attribute and/or any arbitrary (qualified)
                            attribute to refine the identification of the
                            type.
                    </xs:documentation>
                </xs:annotation>
            </xs:attribute>

            <xs:attribute name="extendedSchema" type="xs:anyURI">
                <xs:annotation>
                    <xs:documentation>
                        An identifier for the schema that the value given
                        by the extended attribute is drawn from.
                    </xs:documentation>
                    <xs:documentation>
                        This attribute is normally ignored if the
                        extendedType attribute is not present.
                    </xs:documentation>
                </xs:annotation>
            </xs:attribute>

            <xs:anyAttribute namespace="##other" />
        </xs:extension>
    </xs:simpleContent>
</xs:complexType>

<!--
  -  this definition is taken from the VOTable arrayDEF type
  -->
<xs:simpleType  name="ArrayShape">
    <xs:annotation>
      <xs:documentation>
          An expression of a the shape of a multi-dimensional array
          of the form LxNxM... where each value between gives the
          integer length of the array along a dimension.  An
          asterisk (*) as the last dimension of the shape indicates
          that the length of the last axis is variable or
          undetermined.
      </xs:documentation>
    </xs:annotation>

  <xs:restriction base="xs:token">
    <xs:pattern  value="([0-9]+x)*[0-9]*[*]?"/>
  </xs:restriction>
</xs:simpleType>

<xs:complexType name="SimpleDataType">
    <xs:annotation>
        <xs:documentation>
            a data type restricted to a small set of names which is
            imprecise as to the format of the individual values.
        </xs:documentation>
        <xs:documentation>
            This set is intended for describing simple input parameters to
            a service or function.
        </xs:documentation>
    </xs:annotation>
    <xs:simpleContent>
      <xs:restriction base="vs:DataType">
          <xs:enumeration value="integer"/>
          <xs:enumeration value="real"/>
          <xs:enumeration value="complex"/>
          <xs:enumeration value="boolean"/>
          <xs:enumeration value="char"/>
          <xs:enumeration value="string"/>

          <xs:attribute name="arraysize" type="vs:ArrayShape" default="1"/>
```

```
                    <xs:attribute name="delim" type="xs:string" default=" "/>
                    <xs:attribute name="extendedType" type="xs:string"/>
                    <xs:attribute name="extendedSchema" type="xs:anyURI"/>
                    <xs:anyAttribute namespace="##other" />
                </xs:restriction>
            </xs:simpleContent>
        </xs:complexType>

        <xs:complexType name="TableDataType" abstract="true">
            <xs:annotation>
                <xs:documentation>
                    an abstract parent for a class of data types that can be
                    used to specify the data type of a table column.
                </xs:documentation>
            </xs:annotation>
            <xs:simpleContent>
                <xs:extension base="vs:DataType"/>
            </xs:simpleContent>
        </xs:complexType>

        <xs:complexType name="VOTableType">
            <xs:annotation>
                <xs:documentation>
                    a data type supported explicitly by the VOTable format
                </xs:documentation>
            </xs:annotation>
            <xs:simpleContent>
                <xs:restriction base="vs:TableDataType">
                    <xs:enumeration value="boolean"/>
                    <xs:enumeration value="bit"/>
                    <xs:enumeration value="unsignedByte"/>
                    <xs:enumeration value="short"/>
                    <xs:enumeration value="int"/>
                    <xs:enumeration value="long"/>
                    <xs:enumeration value="char"/>
                    <xs:enumeration value="unicodeChar"/>
                    <xs:enumeration value="float"/>
                    <xs:enumeration value="double"/>
                    <xs:enumeration value="floatComplex"/>
                    <xs:enumeration value="doubleComplex"/>

                    <xs:attribute name="arraysize" type="vs:ArrayShape" default="1"/>
                    <xs:attribute name="delim" type="xs:string" default=" "/>
                    <xs:attribute name="extendedType" type="xs:string"/>
                    <xs:attribute name="extendedSchema" type="xs:anyURI"/>
                    <xs:anyAttribute namespace="##other" />
                </xs:restriction>
            </xs:simpleContent>
        </xs:complexType>

        <xs:complexType name="TAPDataType" abstract="true">
            <xs:annotation>
                <xs:documentation>
                    an abstract parent for the specific data types supported
                    by the Table Access Protocol.
                </xs:documentation>
            </xs:annotation>
            <xs:simpleContent>
                <xs:extension base="vs:TableDataType">
                    <xs:attribute name="size" type="xs:positiveInteger">
                        <xs:annotation>
                            <xs:documentation>
                                the length of the fixed-length value
                            </xs:documentation>
                            <xs:documentation>
                                This corresponds to the size Column attribute in
```

```
                            the TAP_SCHEMA and can be used with data types
                            that are defined with a length (CHAR, BINARY).
                        </xs:documentation>
                    </xs:annotation>
                </xs:attribute>
            </xs:extension>
        </xs:simpleContent>
    </xs:complexType>

    <xs:complexType name="TAPType">
        <xs:annotation>
            <xs:documentation>
                a data type supported explicitly by the Table Access
                Protocol (v1.0).
            </xs:documentation>
        </xs:annotation>
        <xs:simpleContent>
            <xs:restriction base="vs:TAPDataType">
                <xs:enumeration value="BOOLEAN"/>
                <xs:enumeration value="SMALLINT"/>
                <xs:enumeration value="INTEGER"/>
                <xs:enumeration value="BIGINT"/>
                <xs:enumeration value="REAL"/>
                <xs:enumeration value="DOUBLE"/>
                <xs:enumeration value="TIMESTAMP"/>
                <xs:enumeration value="CHAR"/>
                <xs:enumeration value="VARCHAR"/>
                <xs:enumeration value="BINARY"/>
                <xs:enumeration value="VARBINARY"/>
                <xs:enumeration value="POINT"/>
                <xs:enumeration value="REGION"/>
                <xs:enumeration value="CLOB"/>
                <xs:enumeration value="BLOB"/>

                <xs:attribute name="arraysize" type="vs:ArrayShape" default="1"/>
                <xs:attribute name="delim" type="xs:string" default=" "/>
                <xs:attribute name="extendedType" type="xs:string"/>
                <xs:attribute name="extendedSchema" type="xs:anyURI"/>
                <xs:attribute name="size" type="xs:positiveInteger"/>
                <xs:anyAttribute namespace="##other" />
            </xs:restriction>
        </xs:simpleContent>
    </xs:complexType>

    <xs:complexType name="StandardSTC">
        <xs:annotation>
            <xs:documentation>
                a description of standard space-time coordinate systems,
                positions, and regions.
            </xs:documentation>
            <xs:documentation>
                This resource provides a mechanism for registering standard
                coordinate systems which other resources may reference as
                part of a coverage descripiton.  In particular, coverage
                descriptions will refer to components of the STC
                descriptions in this resource via an IVOA identifier.  It
                is intended that an application consuming such coverage
                descriptions be able to semantically interpret the
                identifier without resolving it.  For this reason, once a
                standard STC description is registered with this resource
                type, updating the description is strongly discouraged.
            </xs:documentation>
        </xs:annotation>

        <xs:complexContent>
            <xs:extension base="vr:Resource">
```

```
                    <xs:sequence>

                        <xs:element name="stcDefinitions"
                                    type="stc:stcDescriptionType"
                                    minOccurs="1" maxOccurs="unbounded">
                            <xs:annotation>
                                <xs:documentation>
                                  An STC description of coordinate systems,
                                  positions, and/or regions
                                </xs:documentation>
                                <xs:documentation>
                                  Each system, position, and region description
                                  should have a an XML ID assigned to it.
                                </xs:documentation>
                                <xs:documentation>
                                  Because the STC schema sets
                                  elementFormDefault="qualified", it is
                                  recommended that this element specify the STC
                                  default namespace via an xmlns namespace.
                                </xs:documentation>
                            </xs:annotation>
                        </xs:element>

                    </xs:sequence>
                </xs:extension>
            </xs:complexContent>
        </xs:complexType>

        <xs:complexType name="ForeignKey">
            <xs:annotation>
                <xs:documentation>
                  A description of the mapping a foreign key--a set of
                  columns from one table--to columns in another table.
                </xs:documentation>
                <xs:documentation>
                  This definition that the foreign key is being described
                  within the context of the table containing the key.
                </xs:documentation>
            </xs:annotation>

            <xs:sequence>

                <xs:element name="targetTable" type="xs:token">
                    <xs:annotation>
                        <xs:documentation>
                          the fully-qualified name (including catalog and schema, as
                          applicable) of the table that can be joined with the
                          table containing this foreign key.
                        </xs:documentation>
                    </xs:annotation>
                </xs:element>

                <xs:element name="fkColumn" type="vs:FKColumn"
                            minOccurs="1" maxOccurs="unbounded">
                    <xs:annotation>
                        <xs:documentation>
                          a pair of column names, one from this table and one
                          from the target table that should be used to join the
                          tables in a query.
                        </xs:documentation>
                    </xs:annotation>
                </xs:element>

                <xs:element name="description" type="xs:token" minOccurs="0">
                    <xs:annotation>
                        <xs:documentation>
```

```
                    a free-text description of what this key points to
                    and what the relationship means.
                </xs:documentation>
            </xs:annotation>
        </xs:element>

        <xs:element name="utype" type="xs:token" minOccurs="0">
            <xs:annotation>
                <xs:documentation>
                    an identifier for a concept in a data model that
                    the association enabled by this key represents.
                </xs:documentation>
                <xs:documentation>
                    The format defined in the VOTable standard is highly
                    recommended.
                </xs:documentation>
            </xs:annotation>
        </xs:element>

    </xs:sequence>
</xs:complexType>

<xs:complexType name="FKColumn">
    <xs:annotation>
        <xs:documentation>
          A pair of columns that are used to join two tables.
        </xs:documentation>
        <xs:documentation>
          To do an inner join of data from the two tables, a query should
          include a constraint that sets the value from the first column equal
          to the value in the second column.
        </xs:documentation>
        <xs:documentation>
          This type assumes that it is used in the context of
          implied source (i.e., current) and target tables, as in
          the ForeignKey type's fkColumn.
        </xs:documentation>
    </xs:annotation>

    <xs:sequence>
        <xs:element name="fromColumn" type="xs:token">
            <xs:annotation>
                <xs:documentation>
                  The unqualified name of the column from the current table.
                </xs:documentation>
            </xs:annotation>
        </xs:element>

        <xs:element name="targetColumn" type="xs:token">
            <xs:annotation>
                <xs:documentation>
                  The unqualified name of the column from the target table.
                </xs:documentation>
            </xs:annotation>
        </xs:element>

    </xs:sequence>
</xs:complexType>

</xs:schema>
```

# Appendix B: Compatibility Issues with VODataService 1.0

The working draft version 1.0 of the VODataService schema has been in use in IVOA

registries since about 2008. It is expected that registries will migrate over to version 1.1 gradually and during the transition, there may well be instances of both v1.1 and v1.0 in the same registry. While the metadata structures are the mostly the same (particularly the core VOResource metadata), it is worth enumerating where they are different as this can affect how queries against differing metadata are formed.

- In v1.1, `<schema>` replaces v1.0's `<catalog>`.
- In v1.0, the root element of a table description in a `vs:DataCollection` was `<catalog>`. Consequently, a `<table>` element in a v1.1 record is one level lower than in v1.0.
- In v1.0, the root element of a table description in a `vs:CatalogService` was `<table>`. Consequently, a `<table>` element in a v1.1 record is one level lower than in v1.0.
- Version 1.1's `vs:Coverage` type now contains a `<regionOfRegard>` element. In v1.0, this metadatum was only available via `coverage/stc:STCResourceProfile/stc:AstroCoord/stc:Size`.
- Version 1.1's `vs:TableParam` (for describing table columns) adds `<utype>` and `<flag>` elements. The v1.1 `vs:InputParam` adds a `<utype>` element.

# Appendix C: Change History

**Changes since PR-20100916:**

- updated status for elevation to Recommendation.
- cleaned-up mis-labeled and mis-ordered change history.

**Changes since PR-20100914:**

- added change history for PR-20100412.
- added Note about STC mark-up in 3.2
- reworded sentence describing content of `vs:DataType` in section 3.5.

**Changes since PR-20100412:**

- fix numerous typos discovered in TCG review
- added section 1.1 to describe role of standard in the VO architecture, including diagram.
- corrected frequency range for the UV waveband
- corrected links to reference documents

**Changes since PR-20090903:**

- S3.4: added `<:testQuery>` to `vs:ParamHTTP`
- S3.1.1: in text, added explanation of `vs:Format`
- grammatical clean-up

**Changes since WD-20090508 (v1.10):**

- corrected errors in example in Introduction
- added `<description>` and `<utype>` elements to the `vs:ForeignKey` type for consistency with TAP.
- changed type names `vs:TAP` to `vs:TAPType` and `vs:VOTable` `vs:VOTableType`.